\documentclass{emulateapj}
\pdfoutput=1

\begin{document}

\shortauthors{Berdyugina \& Kuhn}
\shorttitle{Surface Imaging of Proxima b and Other Exoplanets}

\title{Surface Imaging of Proxima b and Other Exoplanets:\\
Topography, Biosignatures, and Artificial Mega-Structures}

\author{S. V. Berdyugina\altaffilmark{1,2}, 
J. R. Kuhn\altaffilmark{2}}

\altaffiltext{1}{Kiepenheuer Institut f\"ur Sonnenphysik, 
Sch\"oneckstrasse 6, D-79104 Freiburg, Germany; sveta@leibniz-kis.de }

\altaffiltext{2}{Institute for Astronomy/Maui, University of Hawaii, 
34 Ohi‘a Ku St., Pukalani, HI 96768, USA; kuhn@ifa.hawaii.edu}

\begin{abstract}
Seeing oceans, continents, quasi-static weather, and other surface 
features on exoplanets may allow us to detect and characterize life outside the solar system.
The Proxima b planet resides within the stellar habitable zone 
allowing for liquid water on its surface, and it may be Earth-like.
However, even the largest planned telescopes will not be able to resolve its surface 
features directly.  
Here, we demonstrate an inversion technique to image indirectly exoplanet surfaces using
observed unresolved reflected  light variations over 
the course of the exoplanets orbital and axial rotation:
ExoPlanet Surface Imaging (EPSI).
We show that the reflected light curve contains enough information to detect both 
longitudinal and latitudinal structures and to map exoplanet surface features. 
We demonstrate this using examples of Solar system planets and moons as well as simulated planets 
with Earth-like life and artificial megastructures.
We also describe how it is possible to infer the planet and orbit geometry from  light curves.
In particular, we show how albedo maps of Proxima b can be successfully 
reconstructed for tidally locked, resonance, and unlocked axial and orbital rotation. 
Such albedo maps obtained in different wavelength passbands can provide
"photographic" views of distant exoplanets. We estimate the signal-to-noise ratio necessary
for successful inversions and analyse telescope and detector requirements necessary 
for the first surface images of Proxima b and other nearby exoplanets.  
\end{abstract}


\section{Introduction}

Resolving surface features (e.g., continents) or global weather patterns on even 
the nearest exoplanets through {\em direct imaging} requires optical systems many km in size 
-- this will not happen soon. On the other hand, emitted and reflected planet light has been 
detected from exoplanet host stars by different observing techniques in a few cases.
The {\em Spitzer} space telescope measured IR light curves of hot Jupiters and detected 
enhanced emission longitudes ("hot spots") in their atmospheres \citep[e.g.,][]{har06,knu07,cow08,cro10}. 
However, the latitudinal information could not be yielded from these data.
High-precision polarimetry has provided first albedo and color measurements
for a hot Jupiter \citep{berdetal2008,berdetal2011}.
Similarly, CoRoT, {\em Kepler} and {\em Hubble} telescopes achieved sensitivity to detect optical 
and near UV reflected light curves from hot Jupiters \citep[e.g.,][]{corot,kepler1,kepler2,hubble}.

Model solutions for albedo maps and orbital parameters based on reflected 
flux or polarization light curves have been previously demonstrated for exoplanets 
\citep[e.g.,][]{cow09,fluri&berd2010,kf11,fk12,sch16}.
With an independent formalism presented in this paper -- ExoPlanet Surface Imaging (EPSI) -- 
we confirm some of the conclusions of those papers
and elaborate applications for planets with and without clouds, with seasonal variations,
photosynthetic organism colonies and artificial structures
built by advanced civilizations. Multi-wavelength observations of such planets may enable 
detection of primitive and even advanced exolife. 
Unlike \cite{cow09} we can recover not only longitudinal information but also plenty
of latitudinal information on exoplanet surfaces. Also, in contrast to \cite{fk12}, 
our modeling suggests that large telescope apertures are needed to infer surface albedo maps. 
For example, mapping Proxima b requires a telescope with the diameter at least 12\,m. 
This paper focuses on  realistic models and expectations for mapping the nearest exoplanets.

We show that moderate signal-to-noise ratio (SNR) time-series
photometry of an exoplanet's reflected light can yield a 2D surface map 
of its albedo using an inversion technique.
Reflected light photometry (or/and polarimetry) of exoplanets when the planet 
rotation axis is inclined with respect to the orbit plane normal direction
offers the greatest possibility for mapping the exoplanet surface or seeing stable 
atmospheric structure with surprisingly high spatial resolution. 
Our numerical technique follows similar algorithms used for mapping 
starspots \citep{berd1998, berdetal1998, berdetal2002}.

The recent discovery of the nearest exoplanet Proxima b \citep{proxb} is an opportunity
for obtaining the first surface (and/or weather) images of a possibly Earth-like exoplanet.  
Here we  investigate in some detail the orbit-rotation parameter space relevant to Proxima b.
We estimate the necessary SNR of the data and evaluate how it can be achieved
with future telescopes.

The paper is structured as follows.
In Section~\ref{sec:mod}, we describe our direct model and an inversion algorithm.
In Section~\ref{sec:sim}, we present simulations and inversions for exoplanets with 
Earth-like albedo distribution due to ocean and land topography. 
We show that, under some circumstances, there is sufficient information 
in the reflected light curve to resolve an accurate map of the exoplanet surface
on the scale of subcontinents.
We also discuss limitations due to data noise, weather system evolution,
and seasonal variations in the cloud and surface albedo. 
In Section~\ref{sec:solsys}, we present inversions for Solar system 
planets and moons as analogs of exoplanets. We can recover global circulation 
cloud patterns on Jupiter, Neptune, and Venus, as well as surface features on 
the Moon, Mars, Io and Pluto.
In Section~\ref{sec:signat}, we present models for spectral (broad-band) imaging
of exoplanets with Earth-like life 
and demonstrate that photosynthetic biosignatures and surface composition
can be unambigously detected with our technique. We also model hypothetical planets
with global artificial structures which can be detected under certain circumstances.
In Section~\ref{sec:acen}, we present simulations and inversions for the Proxima b
planet assuming tidally locked orbits at the 1:1 and 2:3 resonances. 
In Section~\ref{sec:tel}, we investigate and formulate observational requirements
for the telescope size and scattered light level, in order to obtain spectral images
of Proxima b and possible rocky planets in the nearest to the Sun stellar system 
Alpha Centauri A and B. We find that telescopes in diameter of 12\,m or larger
and with a high level of scattered light suppression 
can achieve the required signal-to-noise ratio in planet brightness 
measurements above the stellar glare background. 
We also estimate how many Earth-size and super-Earth planets in water-based habitable zones
(WHZ) of nearby 3500 AFGKM stars can be imaged with our technique depending on the telescope 
aperture for a given scattered light background.
Finally, In Section~\ref{sec:con}, we summarize our results and conclusions.

\section{Model and Inversion Algorithms}\label{sec:mod}

In this section, we describe the model and the inversion algorithm.
The model includes radiation physics and star-planet geometry. 
The inversion algorithm includes criteria for choosing a unique solution
and for evaluating the quality of the recovered image.  

\begin{figure}
\plotone{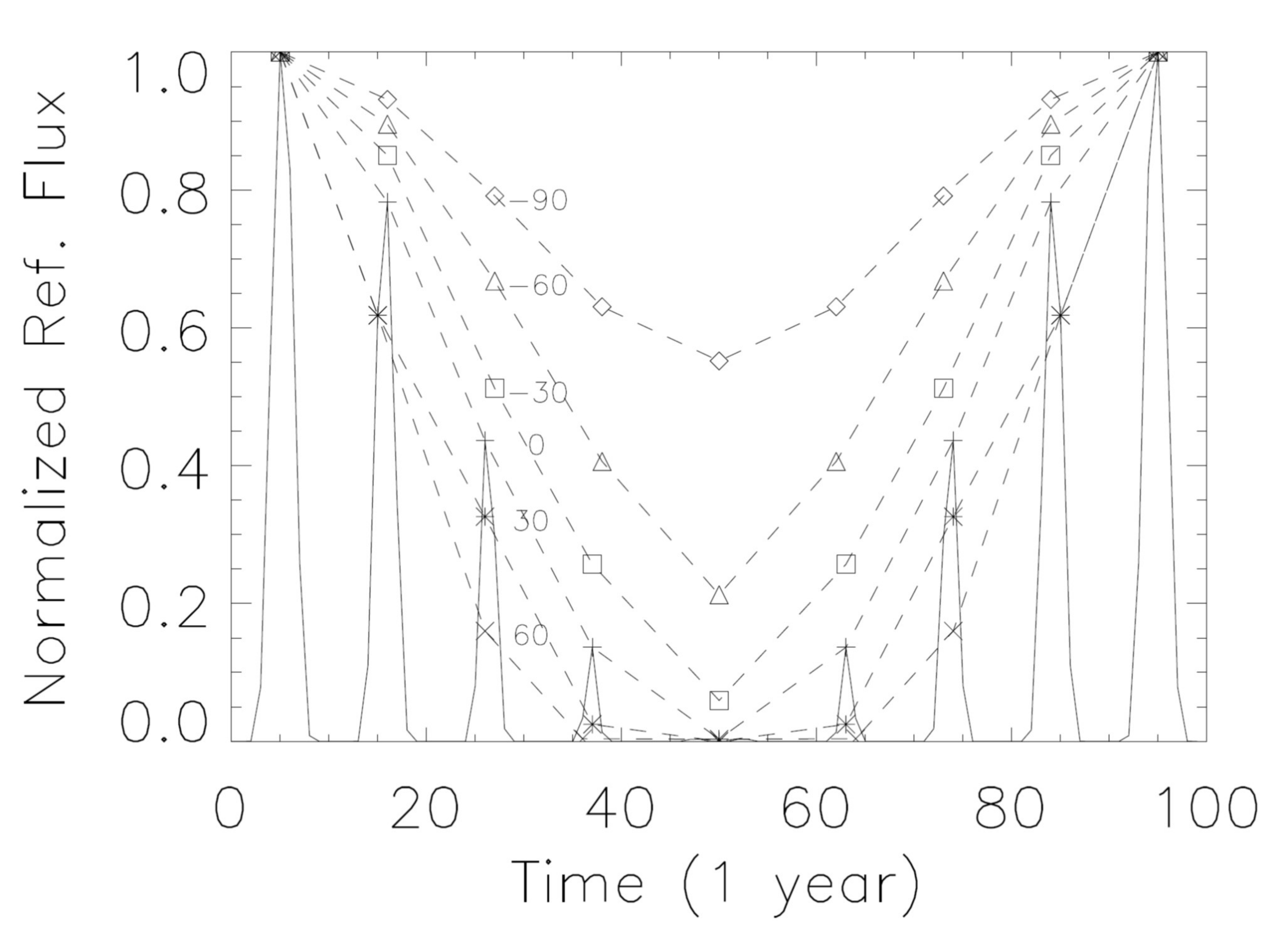}
\centering
\caption{Reflected light curves for an exoplanet with a single high albedo spot 
as a Gaussian with the width about 5 degrees. The planet has an orbit period of 
100 time units with a diurnal period of 10. 
The solid curve shows the light curve for an equatorial spot which goes in and out of
view due to axial rotation and is illuminated fully or in part due to orbital motion.
When the spot is moved to other latitudes, the peak amplitudes are changed.
Symbols joined by dashed lines show the corresponding peak brightness 
for a spot located at the marked planetary latitudes. 
The observer here is located 90$^\circ$ from the planet rotation axis and 
the normal to the plane of the orbit is 45$^\circ$ from the rotation axis. 
These calculations assume a simple $cos\theta$ albedo angle dependence.}
\label{fig:refspot}
\end{figure}

The concept of our EPSI technique is evident if we consider a set of high albedo spots at fixed 
longitude but variable latitude. Figure ~\ref{fig:refspot} shows how the latitude information 
of each reflective spot is encoded in the yearly variation of the peak daily  brightness. 
Each latitude creates a distinct yearly variation profile, while each longitude affects 
the light curve at a distinct daily temporal phase. 
It is evident that de-projecting corresponding daily phased brightness measurements over the course 
of an exoplanet year allows, in principle, reconstruction of the latitude and longitude albedo variation
from orbital and rotational sampling.

In fact, assuming that the surface structure 
does not (or insignificantly) evolve during the time covered by the light curve, 
this information can be inverted into a 2D map of the surface -- 
an indirect imaging technique that is employed in various remote sensing applications,
e.g., asteroid mapping or stellar surface imaging. 
\cite{fk12} employed this approach combined with the Tikhonov
Regularization (TR) inversion algorithm, which minimized discrepances between the model 
and observations by choosing the solution with the minimum albedo gradient, i.e., it
results in the smoothest map solution. 
Here, we describe an algorithm employing the Occamian Approach (OA) inversion technique
\citep{berd1998}. It is based on the Principal Component Analysis (PCA) 
which does not constrain the solution with any prior specific smoothness properties 
(see Section~\ref{sec:inv}). 

\subsection{Direct Modeling}\label{sec:dmod}

\subsubsection{Radiation Physics}\label{sec:dmod_rad}

The reflecting surface of the planet is divided by a grid in longitude and latitude.
Each grid pixel is assigned a geometrical albedo value $A$ for a given wavelength 
or passband and a phase function $P$ defining the angular distribution of the reflected light.
Thus, if $P$ is defined, there are $N=N_{\rm lat}\times N_{\rm lon}$ unknowns, which
comprise an albedo map of the planet with $N_{\rm lat}$ latitudes and $N_{\rm lon}$ longitudes.
This generic description of the surface allows us to compute reflected light 
from various surfaces, including a cloud deck and the planet surface 
with and without an atmosphere. 

Different phase functions can be assigned for different scatterers, e.g., 
Rayleigh for gas and small particles, Mie for cloud droplets, lab measured BRDF 
for various surfaces, etc. \citep[e.g.,][]{berdetal2016,berd2016}.
For simplicity, initial tests presented in this paper were calculated with the
assumption that surface reflection is isotropic, gaseous atmospheric 
scattering is Rayleigh, $\bf{\hat{P}}_{\rm Ray}$, and cloud particles are spherical droplets,
$\bf{\hat{P}}_{\rm Mie}$, if not stated otherwise throughout the paper. 
In some cases, reflected {\em flux} spectra from dense (optically thick) 
high clouds can be described by a composite phase function including 
a Rayleigh scattering contribution from a thin gaseous atmosphere above the cloud deck 
and a diffuse, but wavelength-dependent reflected flux from clouds of a given albedo $A_{\rm c}$. 
Hence, to account for clouds, we assume that each planet map pixel can have contributions 
from clouds and planet surface, with respective filling factors
$f_{\rm c}$ and $1-f_{\rm c}$. The corresponding gaseous atmosphere scattering contributions 
depend on the optical thickness of the atmosphere above the planet surface and above the cloud deck.
Here, we assume that the light scattered by clouds downwards and subsequently reflected 
by the surface is trapped within the atmosphere and does not reach an external observer
(i.e., one cannot "see" through clouds). This approach further simplifies the problem 
and accelerates inversions.

Multiple scattering is computed by solving iteratively 
the radiation transfer equations for intensity and polarization, i.e., for 
the Stokes vector ${\bf I}=(I,Q,U,V)^\mathrm{T}$ 
of a given frequency (omitted for clarity) toward the direction defined by two angles 
($\mu=\cos\theta$, $\psi$):
\begin{equation}\label{eq:rt}
\mu\frac{d\bf{I}(\tau,\mu,\psi)}{d\tau} = \bf{I}(\tau,\mu,\psi) - \bf{S}(\tau,\mu,\psi)
\end{equation}
with the total source function
\begin{equation}\label{eq:sourcef}
\bf{S}(\tau,\mu,\psi) = \frac{\kappa(\tau)\bf{B}(\tau)+\sigma(\tau)\bf{S}_\mathrm{sc}(\tau,\mu,\psi)}{\kappa(\tau)+\sigma(\tau)}
\rm \ ,
\end{equation}
where $\kappa$ and $\sigma$ are absorption and scattering opacities, 
$\bf{S}_\mathrm{sc}$ and $\bf{B}$  are the scattering source function and 
the unpolarized thermal emission, respectively, and $\tau$ is the optical depth in the atmosphere
with $\tau=0$ at the top. The formal solution of Eq.~(\ref{eq:rt}) is \citep[e.g.,][]{Sobolev1956}
\begin{equation}\label{eq:rt_formal}
\begin{tabular}{rcl}
$\bf{I}(\tau,\mu,\psi)$ & $=$ & $\bf{I}(\tau_*,\mu,\psi)e^{-(\tau_*-\tau)/\mu}$ \\
                 & $+$ & $\int_{\tau}^{\tau_*} \bf{S}(\tau',\mu,\psi)e^{-(\tau'-\tau)/\mu}\frac{\mathrm{d}\tau'}{\mu}$ 
				 \rm \ ,
\end{tabular}
\end{equation}
where $\tau_*$ is either the optical depth at the bottom of the atmosphere for the Stokes vector 
$\bf{I}^+(\tau,\mu,\psi)$ coming from the bottom to the top ($\theta < \pi/2$) or the optical depth 
at the top of the atmosphere ($\tau_*=0$) for the Stokes vector $\bf{I}^-(\tau,\mu,\psi)$ coming from 
the top to the bottom ($\theta > \pi/2$).

The scattering source function $\bf{S}_\mathrm{sc}$ is expressed via the scattering phase matrix 
$\bf{\hat{P}}(\mu,\mu';\psi,\psi')$, depending on the directions of the incident ($\mu'$, $\psi'$) 
and scattered ($\mu$, $\psi$) light:
\begin{equation}\label{eq:sourcef_sc}
\bf{S}_\mathrm{sc}(\tau,\mu,\psi) = \int \bf{\hat{P}}(\mu,\mu';\psi,\psi')\bf{I}(\tau,\mu',\psi')\frac{d\Omega'}{4\pi}
\rm \ .
\end{equation}
It has contributions from scattering both incident stellar light and intrinsic thermal emission. 
Their relative contributions depend on the frequency. For instance, for Rayleigh scattering 
the intensity of the thermal emission of an Earth-like planet in the blue part of the spectrum 
is negligible compared to that of the scattered stellar light. The phase matrix 
$\bf{\hat{P}}(\mu,\mu';\psi,\psi')$ is a $4\times4$ matrix with six independent parameters 
for scattering cases 
on particles with a symmetry \citep[e.g.,][]{HansenTravis1974}. In this paper we employ the Rayleigh 
and Mie scattering phase matrices but our formalism is valid for other phase functions too.

The Stokes vector of the light emerging from the planetary atmosphere ${\bf I}(0,\mu,\psi)$ is obtained 
by integrating iteratively Equations (\ref{eq:sourcef}) and (\ref{eq:rt_formal}) for a given vertical 
distribution of the temperature and opacity in a planetary atmosphere. 
We solve this problem under the following assumptions:

	 1) the atmosphere is plane-parallel and static;

	 2) the planet is spherically symmetric;

	 3) stellar radiation can enter the planetary atmosphere from different angles and can be polarized;

	 4) an incoming photon is either absorbed or scattered according to opacities in the atmosphere;

	 5) an absorbed photon does not alter the atmosphere (model atmosphere includes thermodynamics effects 
	      of irradiation);

     6) photons can be scattered multiple times until they escape the atmosphere or absorbed.

These assumptions expand those in \cite{fluri&berd2010}, namely that multiple scattering
on both molecules and particles is accounted for, stellar irradiation can be polarized 
and vary with an incident angle, 
and the planetary atmosphere and surface can be inhomogeneous in both longitude and latitude.
Boundary conditions are defined by stellar irradiation at the top, 
surface reflection at the bottom, and planetary thermal radiation, if necessary. 
Note that the bottom boundary condition refers to reflection from both the planetary surface 
and the cloud deck. 
Depending on the structure of the phase matrix and the boundary conditions, the equations are solved 
for all or fewer Stokes vector components. Normally it takes 3--15 iterations to achieve 
a required accuracy (depending on the surface albedo and optical thickness of the atmosphere).

The stellar light is initially approximated by black body radiation
with the effective temperature of the star, but nothing prevents us to employ real stellar
spectra. Theoretically, the stellar spectrum is removed by normalization to the incident light
(like in the case of the Solar system bodies). In reality, however, stellar spectral lines 
induce irradiation fluctuations and, hence, the signal-to-noise ratio (SNR) variations with wavelength. 
In Section~\ref{sec:tel} we show realistic calculations of the SNR for the nearest exoplanet(s) 
in the Alpha Centauri system in broad photometric bands that is feasible in the near future.

For cloudless Earth-like planets, we used a standard Earth atmosphere \citep{McCl72},
for which we computed Rayleigh scattering opacities and optical thickness depending
in the height in the atmosphere.
For cloudy planets, we assumed an atmosphere of a given optical thickness 
above clouds, while cloud particles scatter light according to the chosen phase function.

Using such atmospheres, we have computed reflected radiation
for a set of incoming and outgoing angles. 
Since the amount of scattered light in a cloudless atmosphere depends on the surface albedo,
we also computed the outgoing radiation for a set of surface albedo values.
These pre-computed sets were used for simulations and inversions of light curves.
It is obvious that the optical thickness of the atmosphere influences the visibility
of the surface. Therefore, for thicker atmospheres there is less information in the
light curve on the surface albedo, and this has to be accounted for while carrying out inversions.
Time-dependent reflected light curves were computed by integrating local Stokes parameters 
over the planetary surface at different rotational and orbital phases. 
Various levels of Poisson (photon) noise were introduced into the theoretical light curves 
to imitate observations. 

\begin{figure}
\plotone{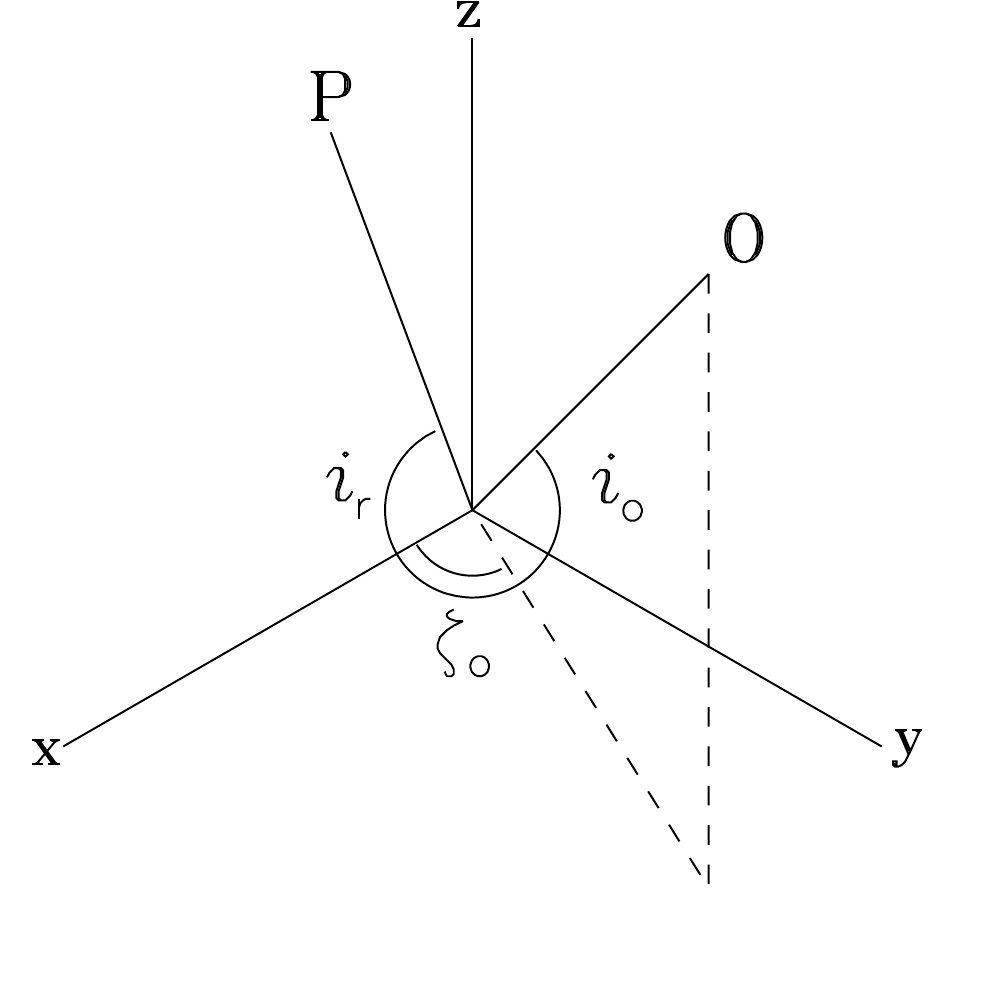}
\caption{Geometry of a planet orbiting a star: $y$ and $z$ axis are in the plane of the sky, 
$x$ axis is toward the observer (perpendicular to the page).
 Planet axis inclination $i_{\rm r}=60^\circ$,
 orbit normal inclination $i_{\rm o}=30^\circ$, and
 orbit normal azimuth $\zeta_{\rm o}=30^\circ$, 
 }
\label{fig:geom}
\end{figure}

\subsubsection{Star and Planet Geometry}\label{sec:dmod_geom}

If the planet is angularly resolved from the star by direct imaging with a large enough telescope, 
its flux can be measured directly. This also implies that the planet flux should exceed background stellar 
scattered light (including telescope scattering and sky background) at the angular separation 
between the planet and the star. 
If the star-planet system is unresolved, the stellar light is added to the planet 
reflected flux. Since the stellar flux is in general variable (e.g., due to magnetic activity 
or oscillations) during the planet orbital period, it is important to have observing tools
to disentangle planetary and stellar variability. In general this is possible because 
1) the starlight spectral and polarization characteristics are different from those of the planet, 
and 2) if the star is resolved from the planet, its variability can be separately measured.

We define the star-planet geometry in a Cartesian frame with the axis $x$ toward observer 
and $y$ and $z$ in the plane of sky (Fig.~\ref{fig:geom}). 
The planet is assumed to rotate around its axis and orbits its star with different
periods, $P_{\rm r}$ and $P_{\rm o}$, respectively. 

The rotational axis of the planet is generally
inclined with respect to the observer with the angle $i_{\rm r}$. Thus, the $xyz$ location
on the planet surface is transformed into $x'y'z'$ 
by rotation by the angle $i_{\rm r}$ toward the observer.
Here, $x'$ is the cosine of the angle between the local normal and the direction to the observer
at a given phase, which defines the visibility function $V$:
the pixel is visible to the observer if $V$ is positive, cf.\ $x'>0$.

The normal direction to the orbit plane is defined by the inclination angle toward the observer 
$i_{\rm o}$ and an azimuth angle $\zeta_{\rm o}$ around axis $z$ (within $xy$-plane). 
To compute the amount of stellar illumination, we are interested in knowing 
the direction toward the star for any planet pixel at any given phase. We assume that the planet
is far enough from the star to neglect the stellar solid angle, i.e., radiation
coming from the star is unidirectional.
Thus, the $(x_{\rm o},y_{\rm o},z_{\rm o})$ direction to the star at a given orbital phase
$\varphi_{\rm o}$ is transformed into $x'_{\rm o}y'_{\rm o}z'_{\rm o}$ by rotations
first by the angle $i_{\rm o}$ and then by the angle $\zeta_{\rm o}$ (Fig.~\ref{fig:geom}).
The cosine between the local normal to the planet surface and the direction toward the star
in the observer reference frame
defines the illumination function $L$: the pixel is illuminated by the star if $L$ is positive, 
i.e.\ $x'x'_{\rm o} + y'y'_{\rm o} + z'z'_{\rm o} > 0$.

The planet reflected flux for given orbital and rotational phases is computed by integrating the intensity 
of the reflected light over the set of visible pixels  weighted by the pixel area, albedo and illumination
fraction. For anisotropic reflection (scattering), it is also weighted by the phase function.

\subsection{Inversion Algorithm}\label{sec:inv}

Recovering a planet surface albedo distribution
from a time series of flux measurements
requires solving an ill-posed problem:
\begin{equation}
D=FS,
\end{equation}
where $D$ is a column vector containing $M$ observations,
$S$ is a column vector containing $N$ pixels on the planetary surface (in longitude and latitude), and
$F$ is the response operator expressed as an $N\times M$ matrix.
The matrix $F$ determines transformation of a planet image into a data or model light curve. 
It includes all the radiation physics and geometry described in the previous section.
Thus our task is to obtain the albedo map $S$ from the data $D$ under the assumptions included into $F$.

A simple inversion of the matrix $F$ could provide an exact solution if $N=M$, but the solution
will be noisy and unstable due to noise in the data. Alternatively, this problem can be solved 
(also for the case $N\ne M$)
by maximizing the overall probability of obtaining the observed data 
(the likelihood function). However, to avoid fitting the data noise, the overall probability
has to be reduced to some "reasonable" level, which leads to multiple possible solutions
for a given probability (likelihood). The choice of the probability density function
to calculate the likelihood function and the solution "goodness" criterion define the inversion
algorithm and the solution itself.

Here we apply an Occamian Approach (OA) algorithm inspired by the "Occam's razor" principle
and having been employed for mapping stellar surface brightness 
from time series of both spectral line profiles and the stellar flux 
\citep[e.g.,][]{berdetal1998,berdetal2002}. In contrast to the broadly used Maximum Entropy (ME) 
and Tikhonov Regularization (TR) methods, the OA algorithm does not define 
an {\it a priori} quality of the solution, such as minimum information (ME) or maximum smoothness
(TR). Instead, the OA chooses a solution $\tilde{S}$ with the information content equal to that
in the data by means of a global maximization of the probability 
to obtain the observed light curve with the minimum number of significant 
principal components \citep[see details in][]{berd1998}. We use the Poisson function 
to compute the likelihood function and the information content in the data.

\subsection{Inversion Quality}\label{sec:iq}

To characterize the overall success of the inversion we define an
Inversion Quality (IQ) parameter as follows.
First, we transform an image $S$ into a "binary" form 
using a threshold albedo, $A_{\rm T}$ (e.g., 0.1): 
$S(A\ge A_{\rm T})=1$ and $S(A<A_{\rm T})=0$.
We do this for both the original ($S$) and recovered ($\tilde{S}$) images.
This operation selects either bright or dark features in the images.
Then, using these transformed images, we compute the IQ parameter per cent as follows:
\begin{equation}
{\rm IQ}=\frac{100\%}{N}\sum_{i=1}^{N}[S_i\tilde{S_i}+(1-S_i)(1-\tilde{S_i})].
\label{eq:iq}
\end{equation}
Thus, IQ is the percentage of pixels in the recovered map having the albedo
range coinciding with that in the original map.
It is obvious that the IQ parameter depends on the chosen albedo
threshold. We found however that for a successful inversion 
IQ variations with the threshold near the center of the albedo value distribution
are not large.
In the following section when discussing the inversion results, we will
use the threshold albedo $A_{\rm T}= 0.1.$ 

Another informative characteristic of the
inversion success is the standard deviation (SD) of the recovered
map with respect to the input map normalized to the maximum albedo
and expressed in percent. This provides an overall measure of the map
goodness.

\begin{figure}
\centering
\includegraphics[width=7cm]{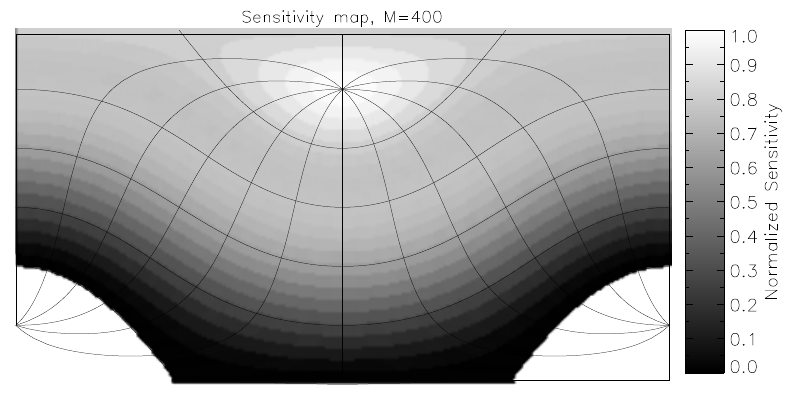}
\includegraphics[width=7cm]{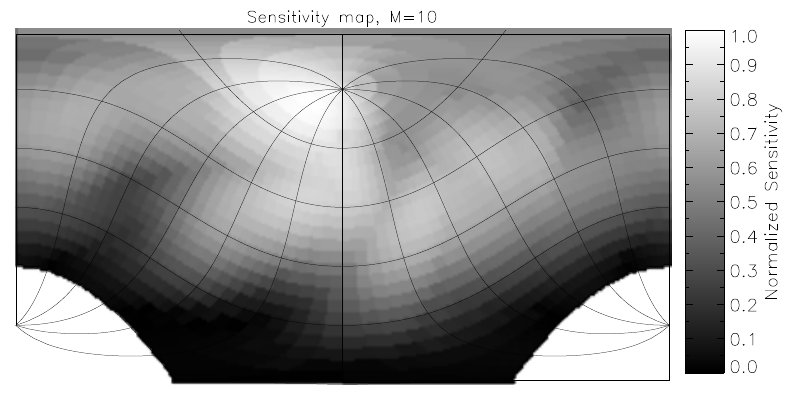}
\caption{Sensitivity maps for model cases with large ($M$=400) and small ($M$=10) amount of data. 
Observed phases are evenly distributed over the orbital period.
The model geometry is the same as in Fig.~\ref{fig:geom}. 
A lack of measurements results in a "spotty" pattern of the sensitivity. Lower latitudes lack
information because of smaller projection area and sparse sampling.}
\label{fig:inv_sen}
\end{figure}

\begin{figure}
\centering
\includegraphics[width=7cm]{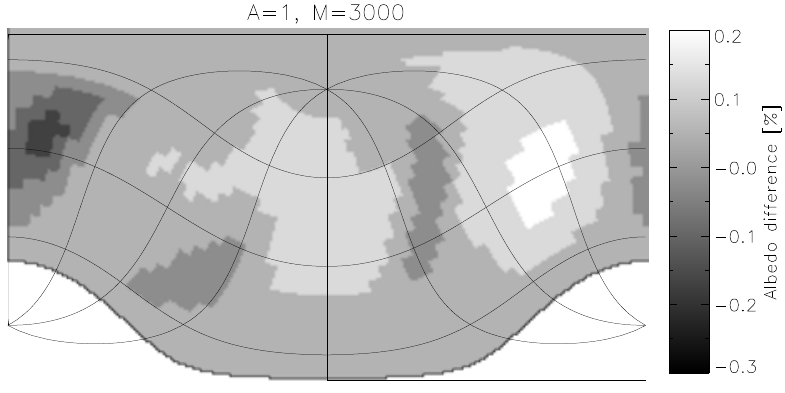}
\caption{Distribution of deviations from the true albedo values $A=1$ for a model with
$i_{\rm r}=60^\circ$, $i_{\rm o}=30^\circ$, and $\zeta_{\rm o}=60^\circ$
shown in Fig.~\ref{fig:geom} and $M$=3000 as in models EN3000 and ES3000 (see Table~\ref{tab:inv}.) }
\label{fig:inv_dev}
\end{figure}

We also compute a sensitivity map in order to evaluate a relative distribution 
of the available information over the planetary surface. It is composed of products 
of the visibility and illumination functions summed over all observed phases:
\begin{equation}
\hat{S_{i}}=\sum_{j=1}^{M} V_{ij} L_{ij} ,
\label{eq:sen}
\end{equation}
where $i=1,N$ runs over all planet pixels.
Thus, $\hat{S_{i}}$ is proportional to the number of photons reflected toward
the observer by the $i$-th pixel during the total observing time, which defines
the error of the restored pixel's albedo. In Fig.~\ref{fig:inv_sen} two
sensitivity maps are shown for model cases with large and small amount of data.
When data sampling is sparse, large gaps in the sensitivity map become apparent.
Since the geometry of the planetary orbit and the rotational axis direction
with respect to the observer
are fixed for a given exoplanet, the sensitivity map helps to optimize
data sampling for the best inversion result. 

One more way to characterize the quality of the recovered map is to
carry out inversion of a light curve for a uniform albedo map (e.g., $A=1$) 
with the same set of the sampled phases and SNR as in the observed light curve. 
The resulting recovered map $\tilde{S}_1$ provides a distribution of expected 
deviations from the true albedo values for each pixel. These include limitations
due to data sampling, visibility and illumination geometry, SNR, and the choice
of the solution "goodness" criterion. Figure~\ref{fig:inv_dev} shows an example
inversion for the $A=1$ map.

\section{Exo-Earth Albedo Inversions}\label{sec:sim}

In this section we present example simulations and inversions for 
an Earth-like planet with continents, ocean and ice caps which allow
for large albedo variations across the planetary surface.
We demonstrate how the information encoded in the reflected light curve 
helps to resolve an accurate map of the exoplanet surface on the scale of continents.
We also discuss limitations due to data noise, weather system evolution,
and seasonal variations in the cloud and surface albedo. 
We characterize inversion results with 
the quality parameters introduced in Section~\ref{sec:iq} and investigate how the quality 
can be maximized. 
We suggest that further improvements in the recovered image quality 
can be achieved by using polarized light curves in addition to the flux
measurements. Such data may be useful for obtaining other local parameters,
like the phase function.

In our tests, we vary the number of planet rotations per orbital period $M_{\rm rot}$
and the number of observed phases per planet rotation $M_{\rm ph}$ as well as SNR, 
geometrical parameters and cloud cover. 
As explained earlier, the total number of measurements is $M=M_{\rm rot}M_{\rm ph}$.

\begin{deluxetable}{lccccc}
\tabletypesize{\scriptsize}
\tablewidth{0pt}
\tablecaption{Test inversion models for an Earth-like planet
\label{tab:inv}}
\tablehead{
\colhead{Model} & 
\colhead{$M_{\rm rot}$} & 
\colhead{$M_{\rm ph }$} &
\colhead{SNR} &
\colhead{IQ} &
\colhead{SD} \\
\colhead{ } & 
\colhead{} & 
\colhead{} &
\colhead{} &
\colhead{[\%]} &
\colhead{[\%]}
}
\startdata   
EN3000   & 60 & 50 & 200 & 89 & 10\\
ES3000   & 60 & 50 & 200 & 89 & 15\\
EN200    & 20 & 10 & 200 & 87 & 11\\
EN400    & 20 & 20 & 20  & 80 & 13
\enddata
\end{deluxetable}

We employ monthly Earth maps produced by the NASA Earth Observatory (NEO) 
satellite missions (https://earthobservatory.nasa.gov/). 
The geometrical albedo maps with no clouds 
are used in Section~\ref{sec:inve_noclouds}.
The effect of seasonal surface albedo variations is analyzed in Section~\ref{sec:inve_seasons}.
To study the effect of variable cloud cover, we analyzed 
a combination of the cloudless albedo maps and monthly cloud cover maps 
weighted by corresponding filling factors in Section~\ref{sec:inve_clouds}.

We have rebinned the maps to the $6^\circ\times6^\circ$ grid,
simulated light curves and carried out inversions as described in Section~\ref{sec:mod}.
The employed models are summarized in Table~\ref{tab:inv}. They are all computed with
$i_{\rm r}=60^\circ$, $i_{\rm o}=30^\circ$, and $\zeta_{\rm o}=60^\circ$.

\subsection{Cloudless Exo-Earths}\label{sec:inve_noclouds}

First, we verify that we can resolve planet surface features in both longitude 
and latitude under favorable conditions, i.e., we assume data with sufficient SNR,
sufficient number of measurements, a favorable (but not unlikely) planet geometry,
optically thin (at a given wavelength) atmosphere, and no clouds.
Inversions for  less favorable scenarios are presented in subsequent subsections.
We use the $6^\circ\times6^\circ$ surface grid for inversions, as it yields
reasonable surface resolution and a comfortable computational effort. This grid has  $N=1800$ 
pixels (when $i_{\rm r}=90^\circ$) which implies that the measurement number $M$ should be
of the same order for a realistic solution. We note that our algorithm also converges 
when $N>M$ but errors increase.

For the first test, we assume that it is possible to obtain planet brightness 
measurements at 50 evenly spaced axial rotational phases with a SNR=200.
The orbital period is assumed to be equal to 60 planetary rotations, resulting
in a total of $M=3000$ measurements.
This case may be representative of an Earth-like planet 
in a WHZ of K-M dwarfs.
As two relevant examples, we carry out inversions for a planet with the terrestrial 
topography visible either from the North or South poles: models EN3000 and ES3000. 
Here we used the albedo map for March 2003, which shows some snow fields 
on the northernmost landmasses and large areas of green vegetation.
 
\begin{figure}
\centering
\includegraphics[width=7cm]{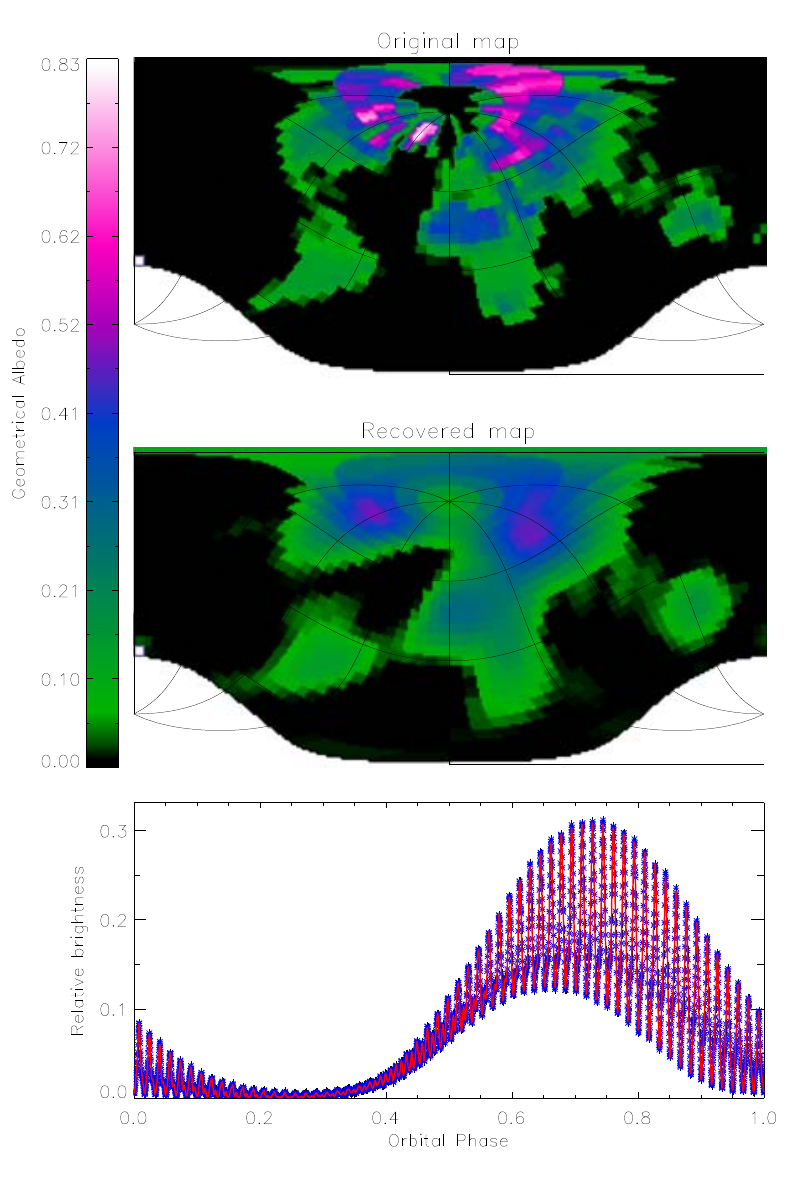}
\caption{Original (top, for March 2003) and recovered albedo map (middle), 
and light curve (bottom) for the model EN3000 (see Table~\ref{tab:inv}). 
The original map is used to simulate the "observed" light curve (blue symbols).
The solid red line light curve is the best fit model corresponding to the recovered
map. Error bars of the simulated data are smaller than the symbol size. }
\label{fig:EN3000}
\end{figure}

\begin{figure}
\centering
\includegraphics[width=7cm]{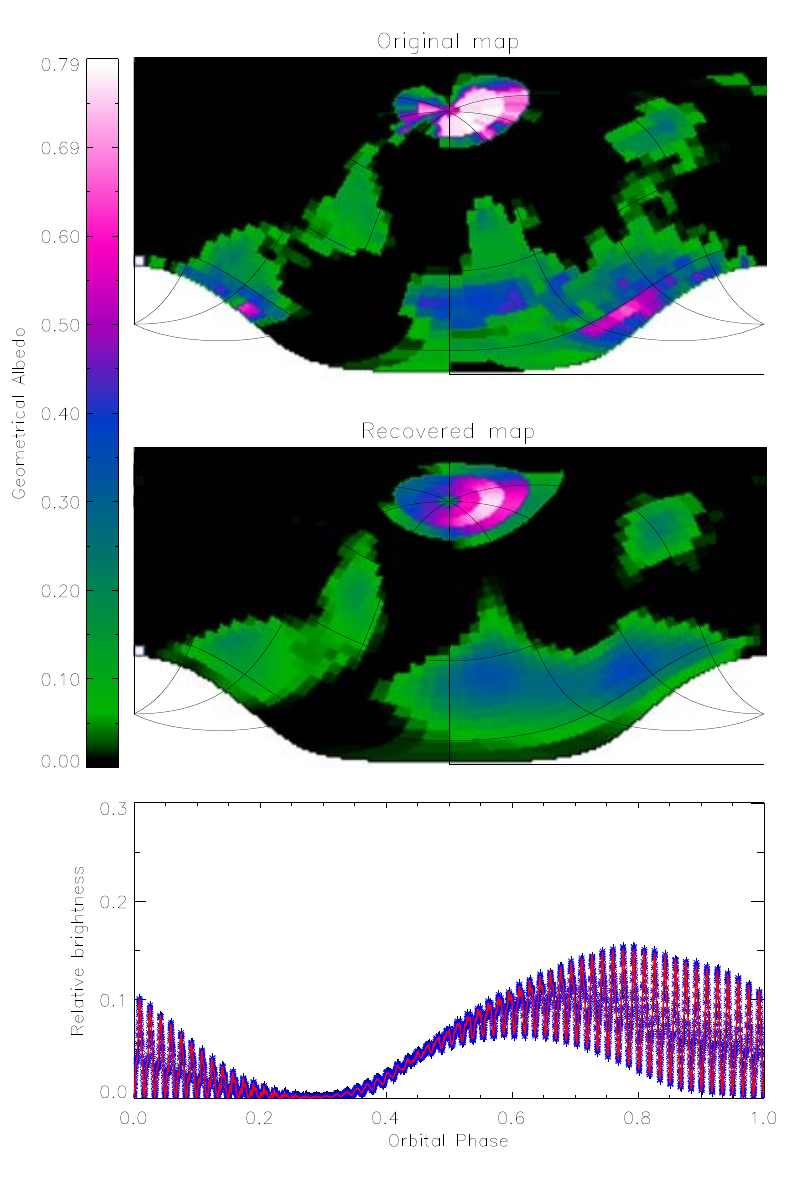}
\caption{The same as in Fig.~\ref{fig:EN3000} for the model ES3000.  }
\label{fig:ES3000}
\end{figure}

The inversion results for these
models are shown in Figs.~\ref{fig:EN3000} and~\ref{fig:ES3000}, respectively.
They successfully recover about 90\%\ of the continental land mass with
$A>0.1$ and the relative standard deviation of 10--15\%.
The shape of the continents and their large-scale albedo features (subcontinents) are
resolved on the scale of Australia and Sahara.
This demonstrates the power of our algorithm.

\begin{figure}
\centering
\includegraphics[width=7cm]{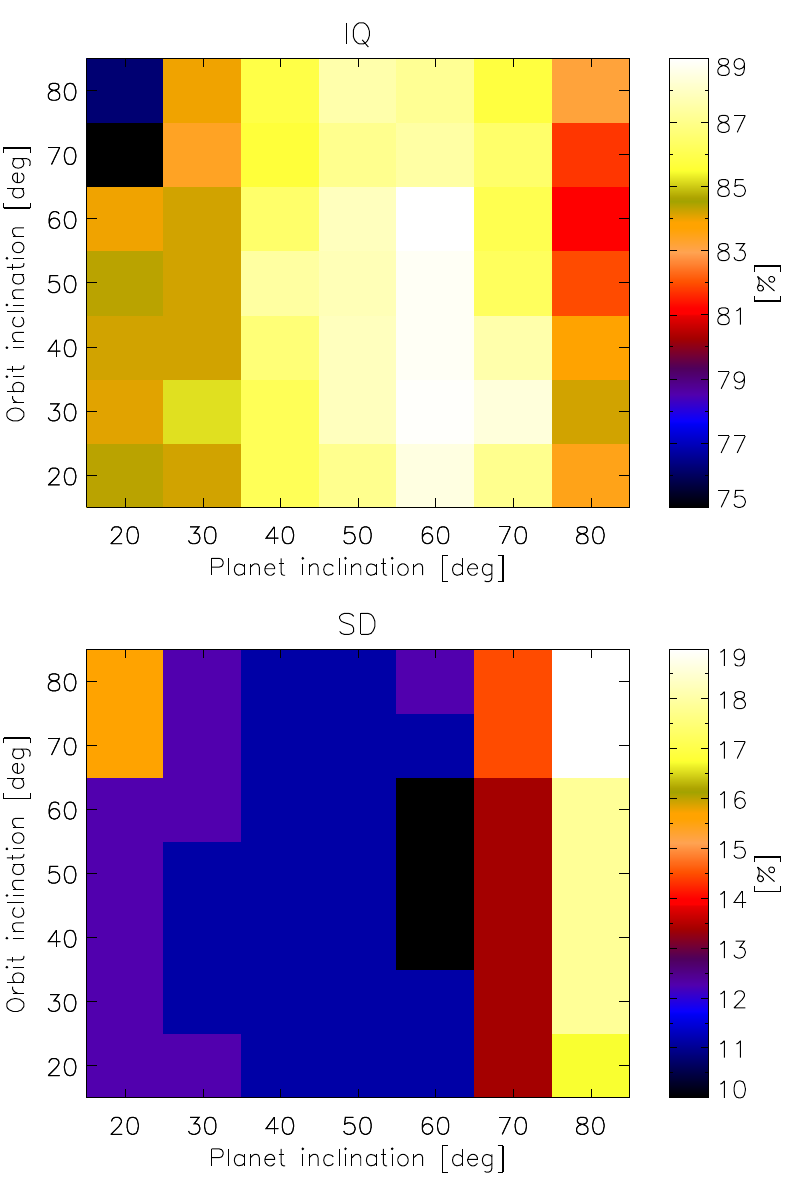}
\caption{Dependence of the IQ parameter (Eq.~\ref{eq:iq}) and map standard deviation SD
on the planet and orbit inclination angles $i_{\rm r}$ and $i_{\rm o}$ for the model EN3000. 
High values are in light tones, and low values are in dark tones.
The median values of IQ and SD are 86\%\ and 12\%.}
\label{fig:iq_inc}
\end{figure}

To investigate the sensitivity of the inversion quality to the geometrical parameters, we have
simulated a grid of models over a range of planet axial and orbital inclinations.
Inversion results for these models are summarized graphically in Fig.~\ref{fig:iq_inc}.
The plots show the dependence of the IQ (Eq.~\ref{eq:iq}) and SD parameters
on the inclination angles. 
Overall our inversion algorithm performs very well at a large range of inclination
angles, except for their extreme values: the median values of IQ and SD are
86\%\ and 12\%, respectively.
At higher planet axis inclination ($\ge80^\circ$), 
it becomes more difficult to distinguish between the upper and lower planet hemispheres.
A combination of low $i_{\rm r}$ with high $i_{\rm o}$ (like Uranus in the Solar system
seen on a near edge-on orbit) is also not favorable for rotational surface imaging.
However, if both $i_{\rm r}$ are $i_{\rm o}$ are low, rotational modulation still produces
enough signal for a reasonable inversion.
There is a particularly favorable combination -- a "sweet spot" -- 
where inversions are most successful due to a good variety of geometrical 
constraints on visibility and illumination of pixels: near 
$i_{\rm r}=50^\circ$ to $70^\circ$ and
$i_{\rm o}$ of $30^\circ$ to $70^\circ$.

\subsection{Seasonal Variations}\label{sec:inve_seasons}

When the planet rotational axis is inclined with respect to the orbital plane,
like for the Earth, one can expect seasonal reflected light fluctuations due to varying
snow, cloud, and biomass surface cover. In this section we analyze 
light curves generated from series of monthly Earth albedo maps, 
but yet without clouds. We consider two ways to evaluate this effect. 

\begin{figure}
\centering
\includegraphics[width=7cm]{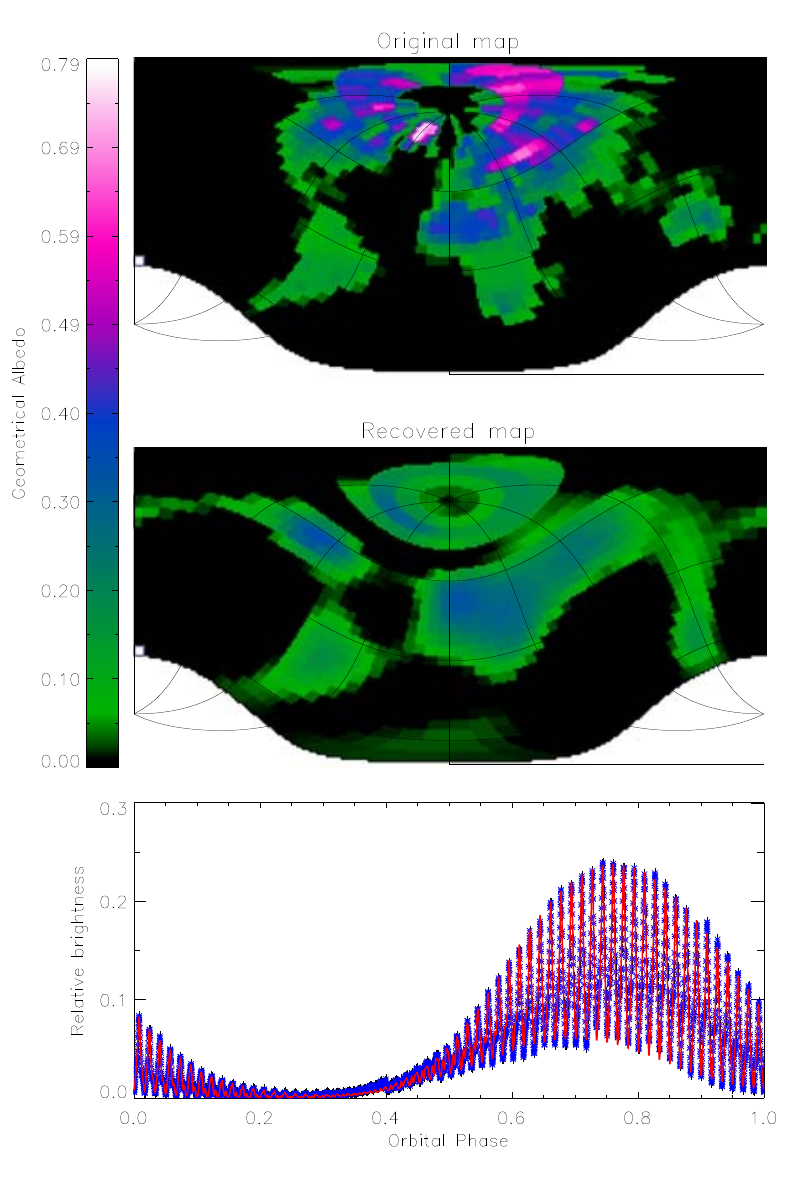}
\caption{Seasonal variation effects in a year-long light curve. 
Top: weighted average of the 12 input albedo maps for the year 2003. 
Middle: recovered average albedo maps. 
Bottom: simulated (crosses) and best-fit (line) light curves.  
Error bars of the simulated data are smaller than the symbol size.
IQ=84\%, SD=11\%. }
\label{fig:EN3000_seasons12m}
\end{figure}

First, we combine 12 subsequent monthly maps (for the year 2003) and carry out inversion 
of a year-long light curve. Hence, seasonal variations are treated as systematic noise, 
and the inversion delivers an average planet albedo map. This map should 
be compared with the average of the 12 input maps weighted by 
the illumination fraction. Such weighting is necessary because the solution will
be dominated by maps seen near the maximum illumination phase (i.e., near the
maximum brightness of the light curve). The result for this test is shown in
Fig.~\ref{fig:EN3000_seasons12m}, with the same planet parameters as for the EN3000 model.
As expected, the quality of the restored map is reduced because of unaccounted systematic
differences in the simulated and best-fit light curves. 

\begin{figure*}
\centering
\includegraphics[width=7cm]{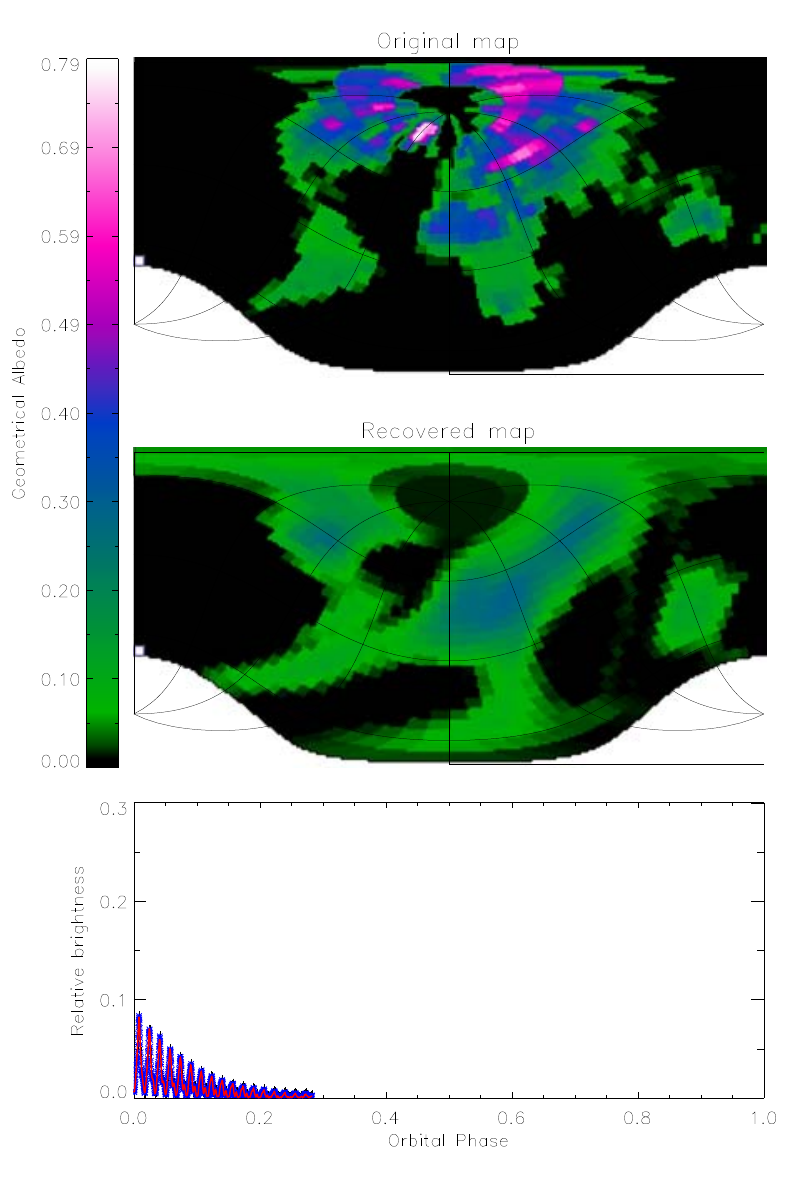}
\includegraphics[width=7cm]{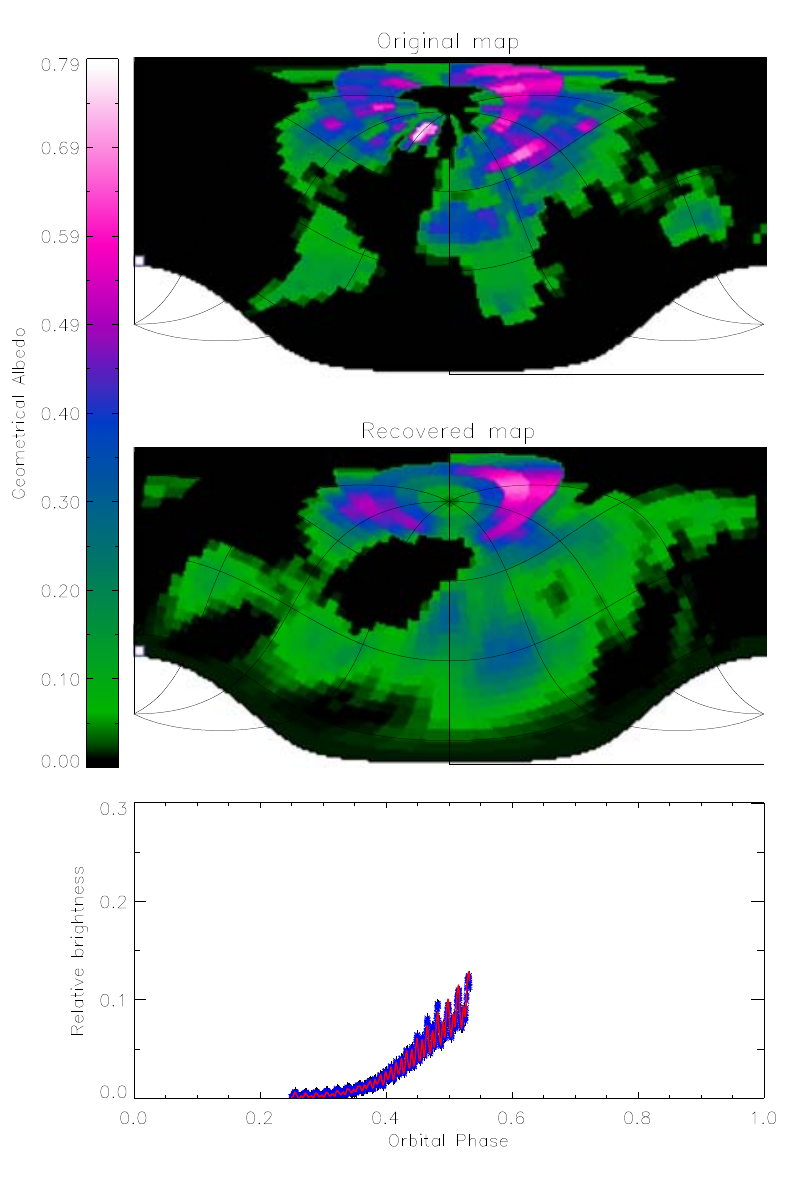}
\includegraphics[width=7cm]{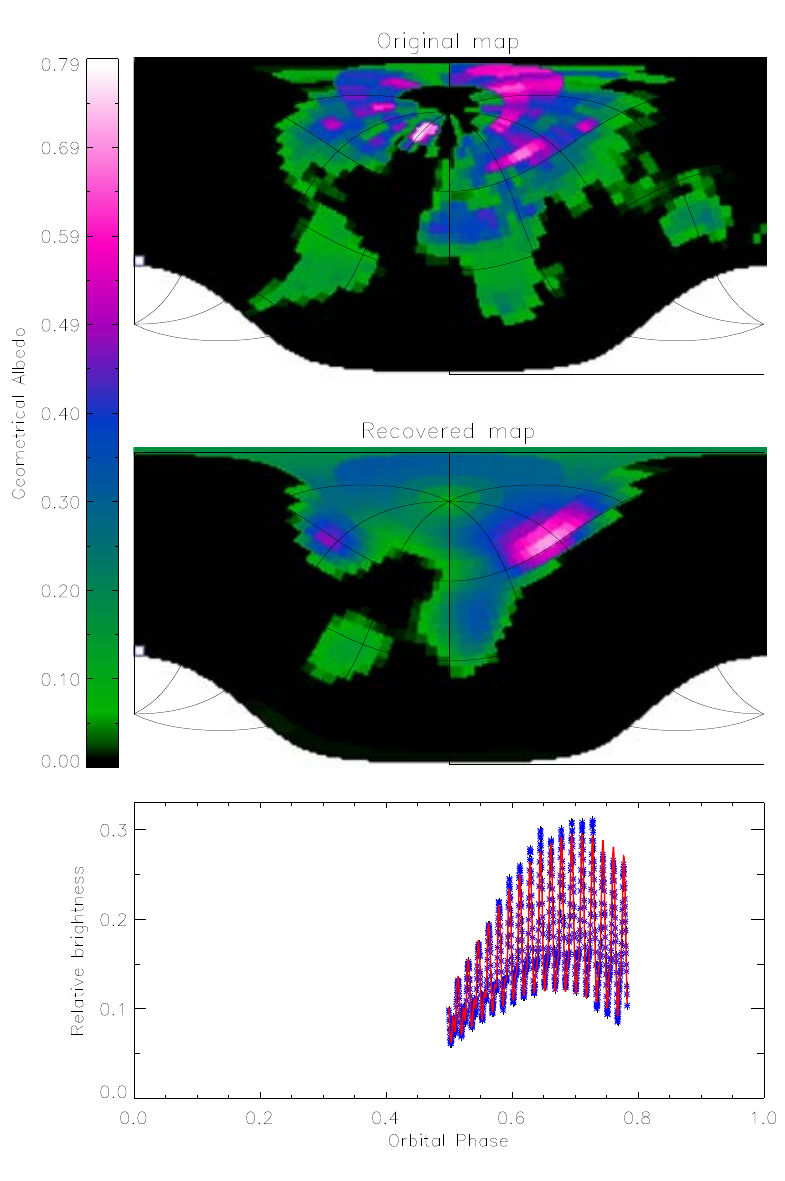}
\includegraphics[width=7cm]{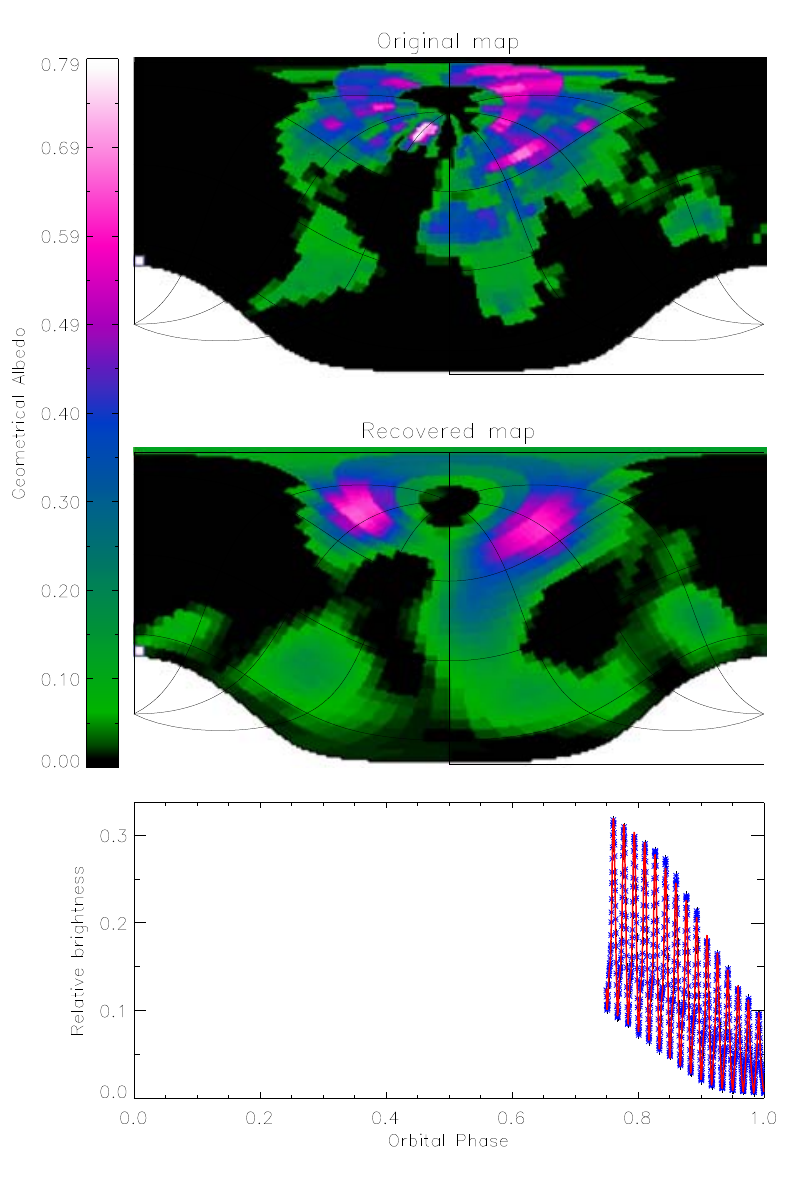}
\caption{The same as Fig.~\ref{fig:EN3000_seasons12m} but for 3-month light curves. 
Top left: months 1-3, IQ=83\%, SD=17\%.
Top right: months 4-6, IQ=82\%, SD=12\%.
Bottom left: months 7-9, IQ=84\%, SD=13\%.
Bottom right: months 10-12, IQ=89\%, SD=13\%.
}
\label{fig:EN3000_seasons_m1-3}
\end{figure*}

Second test is to use the same simulated input as in the first test but to carry out
inversions on parts of the light curve. With sufficient amount of measurements,
one can infer maps for different orbital phases and investigate season progression
on the planet. If such variations are found, they may help to constrain the inclination
of the rotational axis of the planet. The results of such tests are shown in 
Fig.~\ref{fig:EN3000_seasons_m1-3}. 
The quality is obviously reduced as compared to the EN3000 model because of the smaller
amount of data. In addition, only parts of the planet surface could be inferred 
because of limiting illumination at particular orbital phases. 
However, this exercise demonstrates that useful information on the planet surface structures 
can be obtained even from partial light-curve inversions.

\subsection{Partially Cloudy Exo-Earths}\label{sec:inve_clouds}

Planets with liquid water on the surface will most probably experience a water cycle
as on the Earth. This implies that clouds will be forming seasonally and sporadically,
creating various weather conditions. To investigate how the information content in the
recovered exoplanet albedo map depends on the cloud cover, we have computed several
cases with various cloud coverages.

\begin{figure*}
\centering
\includegraphics[width=7cm]{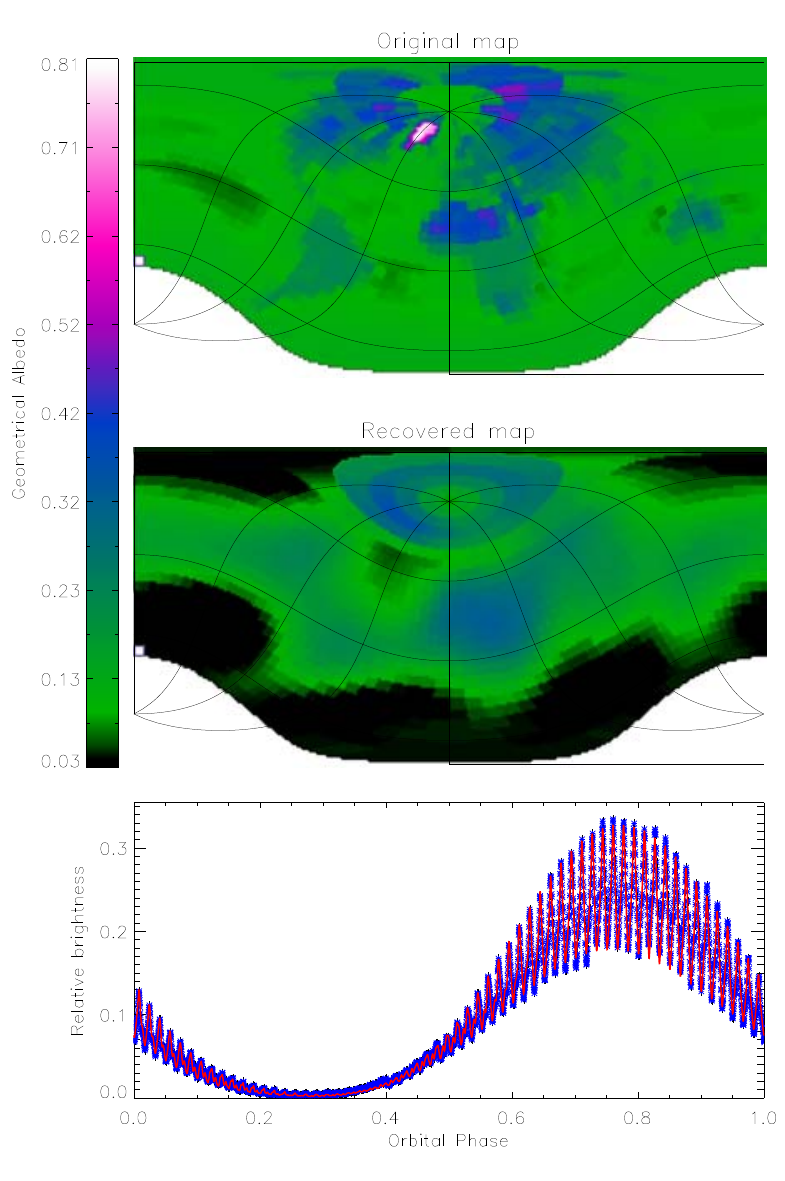}
\includegraphics[width=7cm]{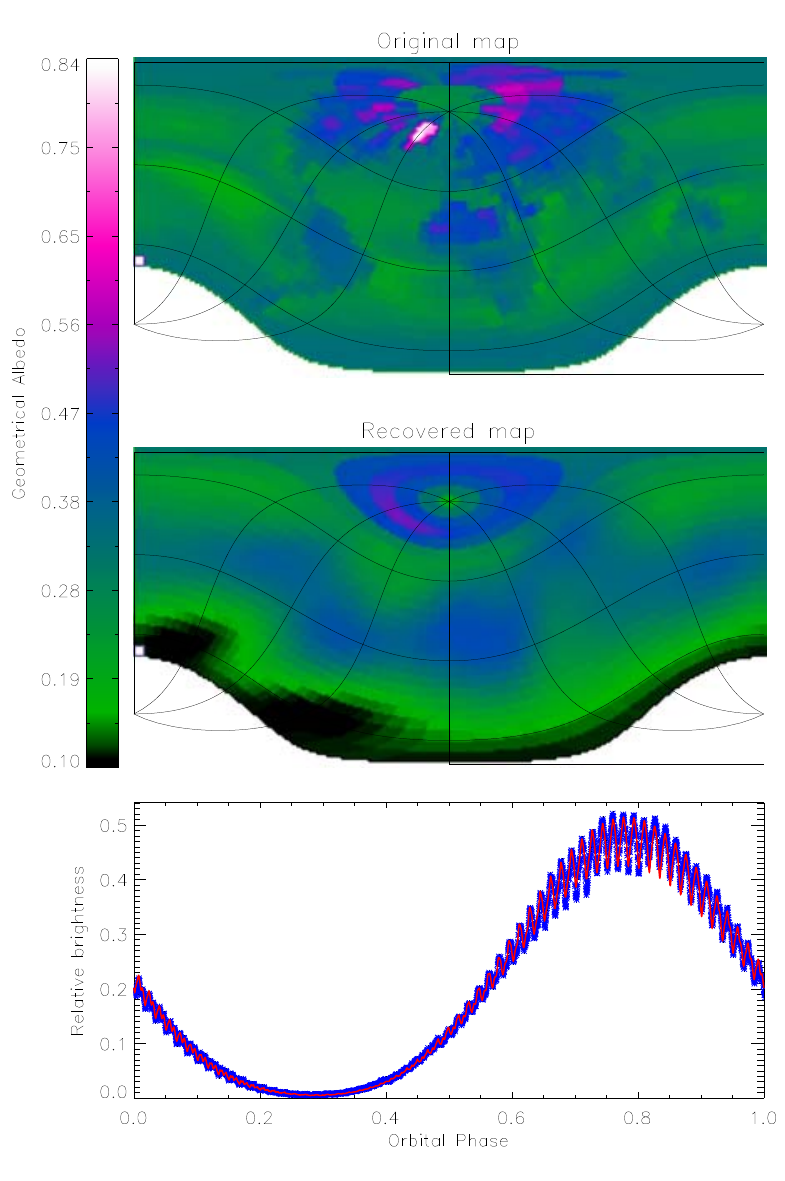}
\caption{The same as Fig.~\ref{fig:EN3000_seasons12m} but for the input maps including contributions from 
monthly cloud coverage maps. The global cloud cover has been reduced (as compared to the measurements 
on the Earth) by the factor 0.1 (left, IQ=74\%, SD=10\%) and 0.3 (right, IQ=78\%, SD=13\%).  
}
\label{fig:EN3000_clouds_m1-12}
\end{figure*}

As in Section~\ref{sec:inve_seasons}, we have combined 12 subsequent monthly maps for the year 2003
and augmented them with corresponding monthly cloud cover maps (also from the NEO website). 
The cloud maps were considered as masks with each pixel being the cloud filling factor $f_{\rm c}$.
The filling factor of the cloudless surface was correspondingly reduced by the factor $1-f_{\rm c}$.
In addition, we have scaled the cloud cover by a global factor in order to quantify how optical 
thickness of clouds affects our ability to see the planet surface. 
As indicated above, the recovered map is to be compared with the average of the 12 input maps
(with seasonal variations of both the surface albedo and cloud cover) weighted by 
the illumination fraction. Two test results are shown in
Fig.~\ref{fig:EN3000_clouds_m1-12}, with the same planet parameters as for the EN3000 model
and global scaling of cloud coverage by factors of 0.1 and 0.3. 
As expected, the quality of the restored planet surface albedo maps is significantly reduced 
for heavily clouded planets. However, longer series of measurements should reveal high-contrast
surface features in more detail.

If clouds dominate the planet albedo, 
the cloud distribution and thickness can be detected on such planets. In this case,
rotational variation amplitude reduces, and  
the cloud light curve signal approaches that of a homogeneous sphere, as shown in
Fig.~\ref{fig:inv_dev}. 
The recovered cloud distribution and thickness can provide additional constraints 
on the exoplanet climate and heat circulation in its atmosphere.
The cloud physical and chemical composition can be retrieved with the help of polarization
measurements as discussed in Section~\ref{sec:pol}.

\subsection{Limited Data Inversions}\label{sec:lim}

The inversion technique is surprisingly forgiving of S/N ratio and number of observation points.
In this section, we investigate how inversions perform under
less favorable conditions.

\begin{figure}
\centering
\includegraphics[width=7cm]{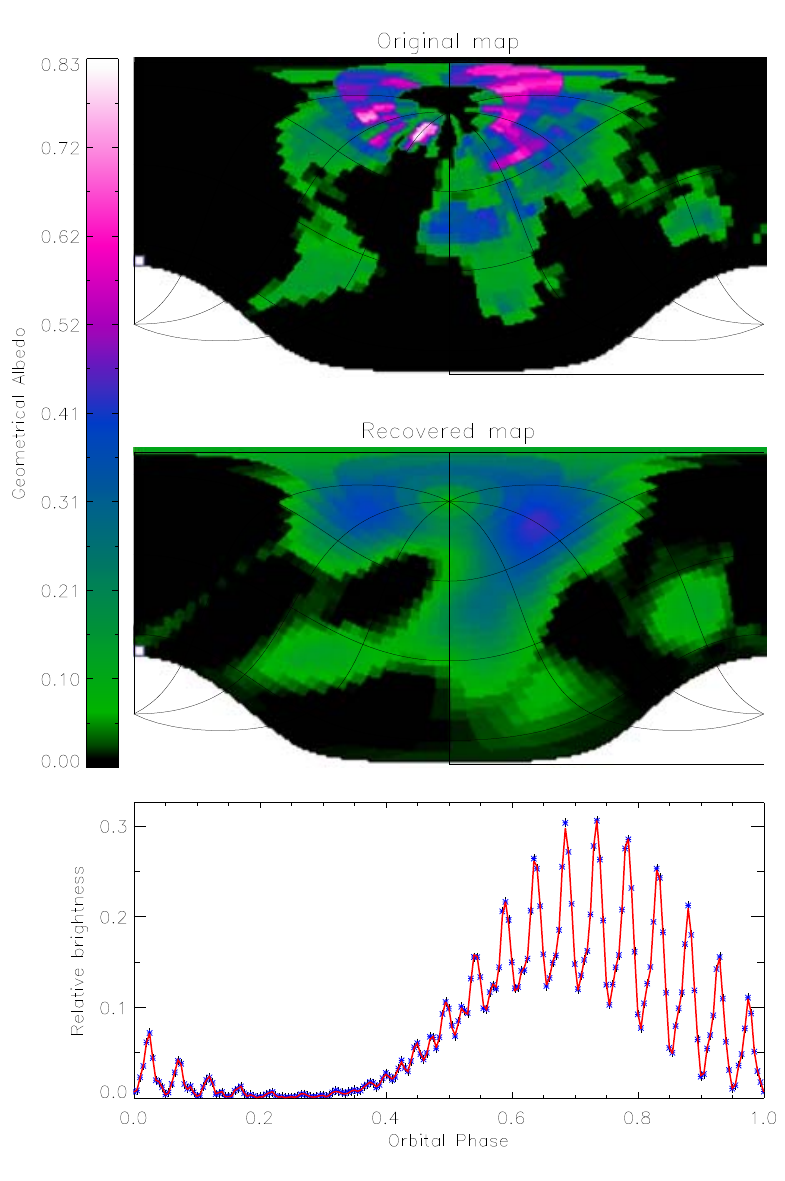}
\caption{The same as in Fig.~\ref{fig:EN3000} but for the model EN200: 
$M_{\rm rot}=10$, $M_{\rm ph}=20$, SNR=200, IQ=87\%, SD=11\%.}
\label{fig:EN200}
\end{figure}

Here we compute the inversion with a reduced number of observations per planet rotation period, such as: 
$M_{\rm rot}$ is reduced to 20 and 
$M_{\rm ph}$ is reduced to 10, and vice versa, when 
$M_{\rm rot}$=10
$M_{\rm ph}=20$.
Hence, the total amount of measurements in both cases is 200, well below $N$
(model EN200 in Table~\ref{tab:inv}).
The geometrical parameters are the same as before, and the orinal map
is for March 2003.
Despite this difference in $M_{\rm rot}$ and $M_{\rm ph}$, 
these two cases result in maps of similar quality.
In Fig.~\ref{fig:EN200} we show one of these cases.
It is only somewhat poorer as compared to the "ideal" inversion 
presented in Fig.~\ref{fig:EN3000}. The major part of the landmass
is recovered with realistic albedo contrast  (but without some details) 
with IQ=87\%\ and SD=11\%.
We conclude that it is the total number of measurements, not necessarily 
the number of measurements during a single exoplanet day that encodes the
latitudinal information. 

\begin{figure}
\centering
\includegraphics[width=7cm]{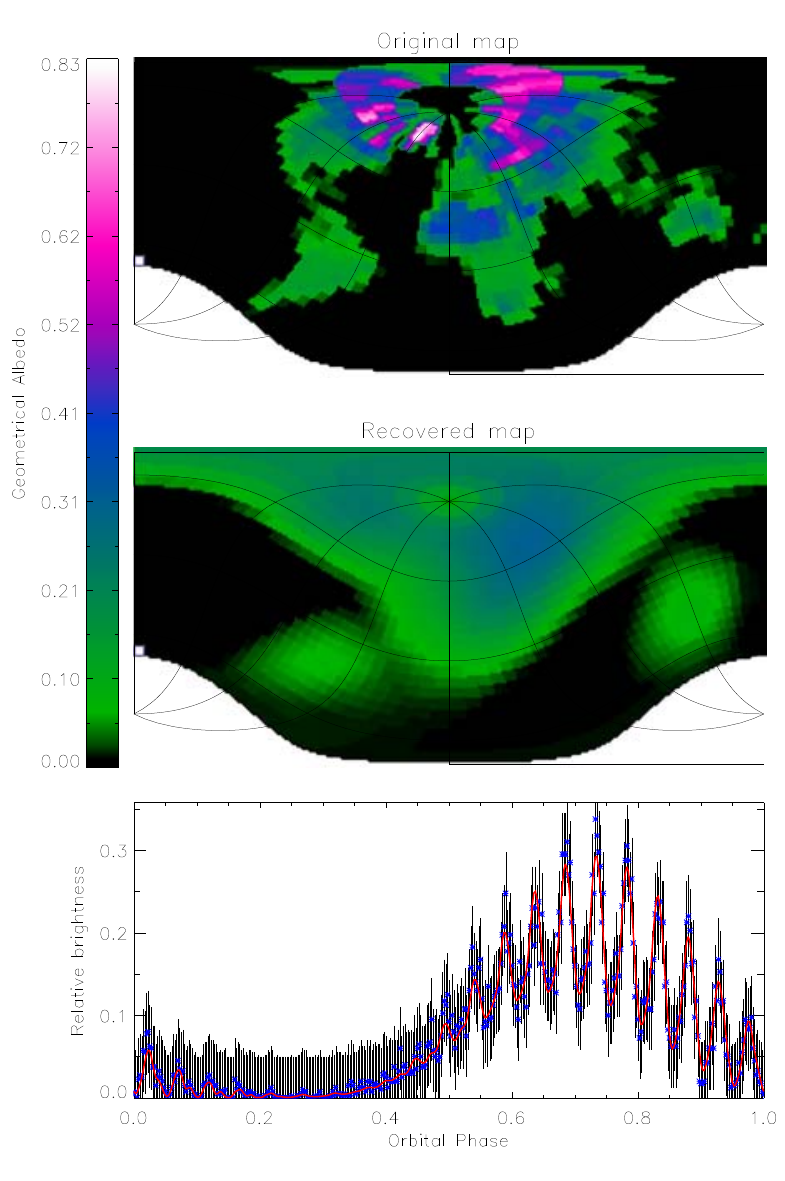}
\caption{The same as in Fig.~\ref{fig:EN3000} but for the model EN400: 
 $M_{\rm rot}=M_{\rm ph}=20$, SNR=20, IQ=80\%, SD=13\%.
Error bars of the simulated data are shown with vertical lines. }
\label{fig:EN400}
\end{figure}

It is also interesting to see how the measurement S/N affects the output images.
The case for $M_{\rm rot}=M_{\rm ph}=20$ and SNR=20 is shown in Fig.~\ref{fig:EN400}
(model EN400 in Table~\ref{tab:inv}).
Here, the fraction of the recovered continent land mass 
is reduced to 80\%\ and SD to 13\%, and many details are missing.
However, even with this low SNR, there is a high-contrast spatial structure 
that we could interpret as evidence of a continental water world.

\subsection{Geometrical Parameter Retrieval}\label{sec:retr}

The light curve contains more information than just an albedo map or cloud cover. 
It strongly depends on the geometrical parameters of the system. 
Therefore, it is possible to retrieve the albedo map as well as the orbital
and planetary axis angles  $i_{\rm o}$, $\zeta_{\rm o}$ and $i_{\rm r}$.

\begin{figure}
\centering
\includegraphics[width=7cm]{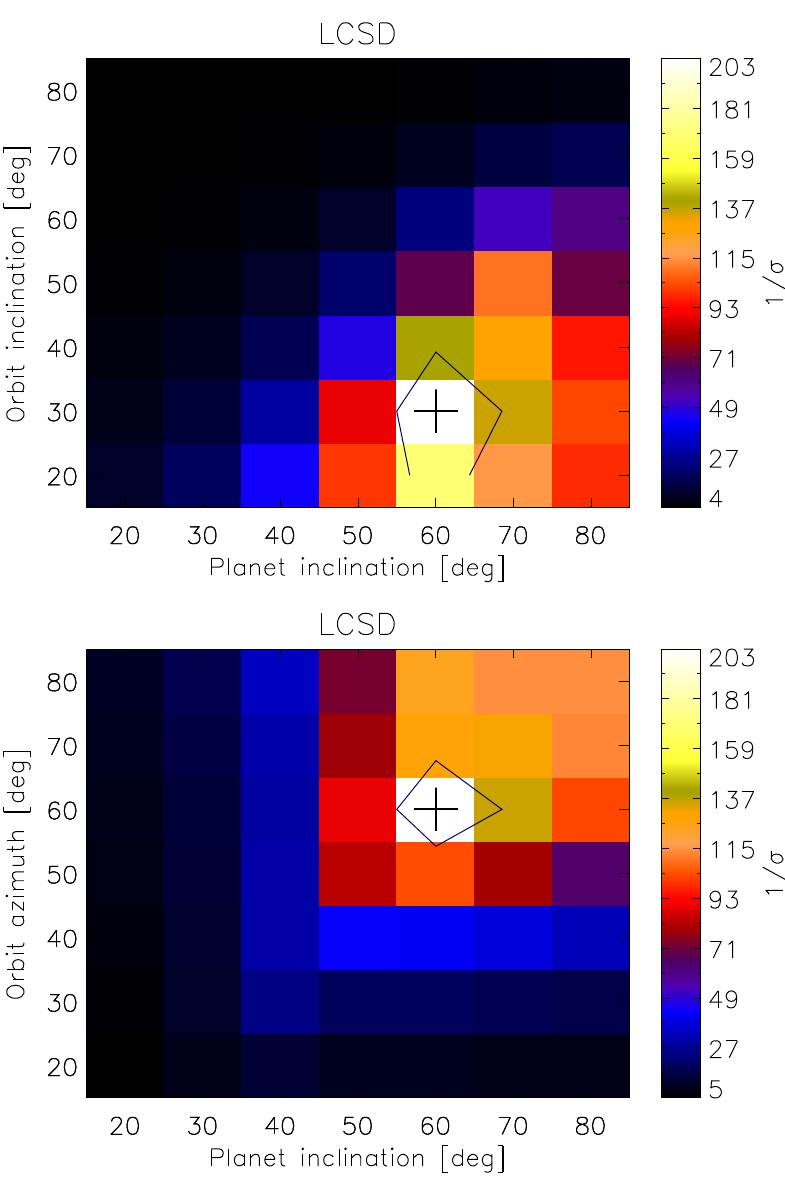}
\caption{Retrievals of the angles  $i_{\rm o}$ and $i_{\rm r}$} for the model EN3000 (SNR=200)
by minimizing the discrepancy between the "observed" and modelled light-curves within the angle domain.
The plots show inverse of light-curve standard deviations (LCSD). 
True values of the angles are marked with crosses. 
Contours show 67\%\ confidence levels. 
\label{fig:retr_sn200}
\end{figure}

Here we retrieve these angles by carrying out inversions and minimizing the discrepancy 
between the "observed" and modelled light-curves within the angle domain. 
We have found that a high SNR light curve (100 or more) allows for a quite accurate 
retrieval of the angles. An example for SNR=200 is shown in Fig.~\ref{fig:retr_sn200}.

\begin{figure}
\centering
\includegraphics[width=7cm]{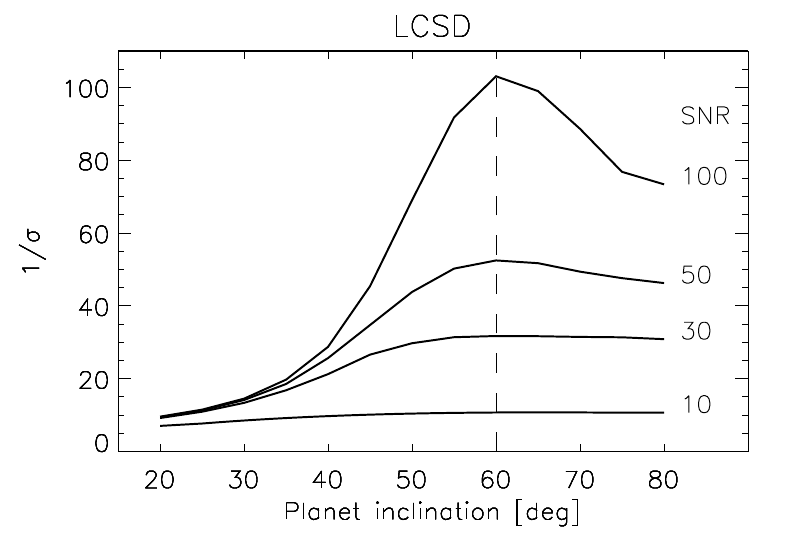}
\caption{Retrievals of $i_{\rm o}$ only for the model EN3000 with SNR=10--100. 
The vertical axis shows inverse of light-curve standard deviations (LCSD). 
The true value is marked with a dashed vertical line.}
\label{fig:retr_sn_i}
\end{figure}

Obviously, noisier data yield weaker constraints on both the albedo map and angles. 
For instance, for SNR $<$ 30, even if the albedo map can be reasonably retrieved,
there is simply not enough information to create a surface map without independent 
information about the planet geometrical parameters.
In fact, both the inclination 
and azimuth  of the orbit can be evaluated from direct imaging data even in such noisier data, 
if the planet is well-resolved from the star. The planet's sky projected position with respect 
to the star changes over the course of its orbit, and the shape and orientation
of this projected trajectory constrain the  $i_{\rm o}$ and $\zeta_{\rm o}$
angles as well as its eccentricity (astrometric orbit). When this information
is employed, the retrieval procedure can successfully converge to the true
planet axis inclination angle even at low SNRs. An example of such a case is
shown in Fig.~\ref{fig:retr_sn_i}.

\subsection{Polarized Reflected Light}\label{sec:pol}

With few assumptions the reflected planetary light will be linearly polarized, 
in some cases at the 10\%\ or larger level.  
From Earth-like planets the light is polarized mostly due to
scattering by molecules (Rayleigh), droplets or particles (clouds) 
in their atmospheres or reflection from the land or ocean surface \citep{earthpol}.
In contrast, the direct stellar light,
instrument and seeing-scattered starlight will have a much smaller linear polarization 
than the angularly resolved exoplanet -- in many cases smaller than $10^{-5}$. 
We will consider linearly polarized light-curve inversions in detail in a future paper, 
but suffice it here to note that there
is considerably more information in the polarized exoplanet light-curves. 

For example, the Stokes Q and U linearly polarized exoplanet light are
sensitive to orbital parameters \citep{fluri&berd2010}, but eliminate 
the residual scattered starlight that generally contaminates the exoplanet photometry. 
Illumination variation due to, for example, elliptical orbits can also be accounted for 
in polarization inversion.  
Also, the mean component of the exoplanet rotationally modulated light polarization 
direction can provide the exoplanet's rotation axis inclination geometry independent 
of details of the surface albedo map.

More interestingly, the exoplanet scattered light polarization angle is sensitive 
to the exoplanetary scatterer's latitude and longitude,
in a way that is completely independent from the corresponding unpolarized flux. 
This provides an additional  constraint on the albedo map that is distinct
from the seasonal and diurnal brightness variations.
 
Finally, reflection from the surface introduces polarized spectral features 
which can help to identify
surface composition \citep{polder} and biosignatures \citep{berdetal2016}. 
To distinguish true surface features from those of clouds, one has to use
polarimetry simultaneously with flux measurements.
In this case, it is possible to determine the scattering phase function,
since polarization is strongly dependent on the scattering angle
and properties of scatterers 
(cloud particles, surface fine structure, vegetation, rocks, etc.).

While leaving the subject
of polarized light curve inversion for another paper, we expect that using
polarized light will improve the sensitivity and allow inversion for
other atmospheric and surface properties.

\section{Solar System Albedo Inversions}\label{sec:solsys}

In this section, we present inversions for some of the Solar system 
planets and moons as analogs of exoplanets. We demonstrate how our EPSI technique can recover 
global circulation cloud patterns as well as solid surface features without a surrounding 
low-albedo ocean. 

For our tests we employ high-resolution composite "photographs" of planets 
and moons delivered by various space missions and telescopes. 
From each such a true-color image we extract three images for red, green and blue 
(RGB) bands. These images are normalized to the planet average visible geometrical 
albedo. Then, corresponding RGB light-curves are simulated and inverted.
A composite image constructed from RGB recovered maps allows then for
studying chemical composition of spatially resolved surface features by 
means of low-resolution spectral analysis.

Our inversions in this section assume $M_{\rm rot}=60$, $M_{\rm ph}=50$, SNR=200, 
$i_{\rm r}=60^\circ$, $i_{\rm o}=30^\circ$, and $\zeta_{\rm o}=60^\circ$.
Only for Jupiter we assume $i_{\rm r}=30^\circ$, so that its famous red spot
is better visible.

\subsection{Planets with Thick Clouds }\label{sec:inve_solsys_clouds}

\begin{figure*}
\centering
\includegraphics[width=17cm]{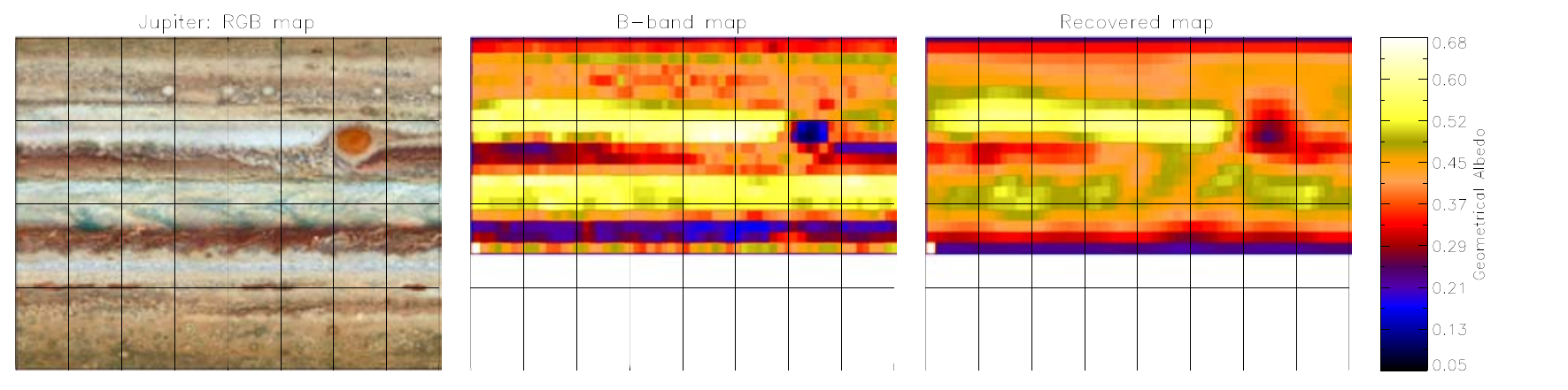}
\includegraphics[width=17cm]{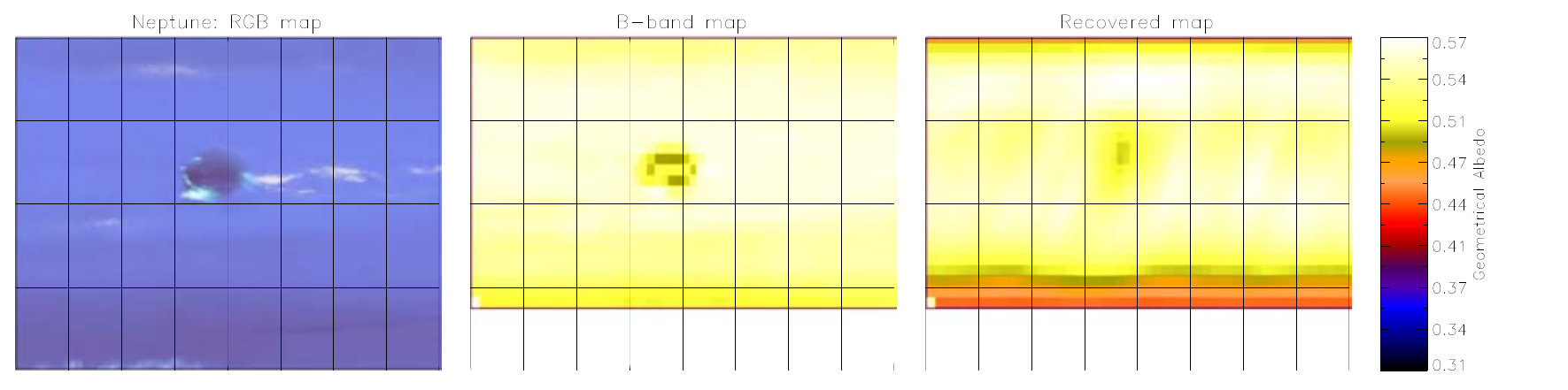}
\includegraphics[width=17cm]{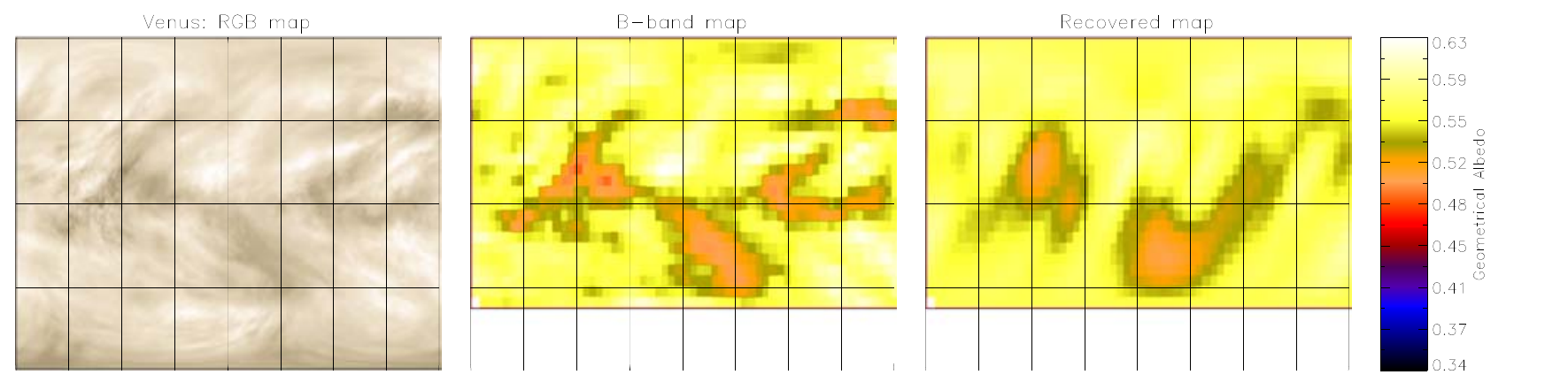}
\caption{Albedo maps for the Solar system planets with thick clouds. 
In the first column, are original RGB maps of higher resolution,
in the second column are original maps rebinned to the $6^\circ\times6^\circ$ grid in a particular band,
and in the third column are recovered maps in the same band. 
The assumed geometrical parameter values are  
$i_{\rm r}=60^\circ$, $i_{\rm o}=30^\circ$, and $\zeta_{\rm o}=30^\circ$,
except for Jupiter $i_{\rm r}=30^\circ$.
The original rebinned maps are used to simulate the "observed" light curve from
which recovered images are obtained using inversions.
}
\label{fig:solsys_cloud}
\end{figure*}

Dense planetary atmospheres have vigorous circulation imprinted in their cloud pattern,
like in zones, belts, vortices, jets, etc., on Jupiter, Saturn, Venus, etc. 
3D models predict similar patterns to exist on exoplanets too
\citep[e.g.,][]{heng2015,amu2016,kom2016}. Hence, images of Solar system planets 
with clouds can be used as a good example for testing our inversion technique 
to map the albedo of planetary atmospheres with a thick cloud deck. 
In this section, we employ images of Jupiter, Neptune, and Venus for such tests. 
The original and recovered images for cloudy planets
are shown in Fig.~\ref{fig:solsys_cloud}. The details for each planet are presented
in the following subsections.

We note that particles in dense clouds are usually larger than the wavelength
of the optical light. Therefore, the Rayleigh approximation is not useful,
and the Mie approximation is applicable only for spherical particles.
In gas giants, such as Jupiter and Neptune, particles are generally
nonspherical (ices) and require complicated phase functions. 
Therefore, the shape of their light curves is also affected by phase functions 
\citep[e.g.,][]{dyu16}.
In this section, for simplicity, when generating flux light curves 
from cloudy Solar system planets and applying inversions to them, 
we assume that clouds scatter isotropically with their geometrical 
albedo to be recovered by the inversion routine at different wavelengths.
Thus, our first goal is to investigate how precisely low-contrast cloud structures
can be recovered by the EPSI technique.

\subsubsection{Jupiter}\label{sec:inve_solsys_clouds_jup}

For Jupiter, we employ a global, high-resolution RGB map constructed using the {\em Hubble}
telescope as a part of the OPAL project \citep{jup_hub}. 
We have extracted from this image a part with bright zones and the Great Red Spot (GRS) 
within the latitudes of $\pm60^\circ$ and longitudes of $\pm90^\circ$.
Then, a full surface map from this part was constructed
by rebinning it to the $6^\circ\times6^\circ$ grid for all longitudes and latitudes.
The Jupiter average visible geometrical albedo of 0.52. Correspondingly, after 
normalization, the average albedos of the individual RGB images are about 0.6, 0.5 and 0.4, 
respectively. These are close to measured values for Jupiter \citep{kar1994}.  

We have used this reduced, normalized, and rebinned RGB image for our inversions
as an example of exoplanets with a global circulation cloud pattern.
Also, we made it visible from the south pole, so that the GRS is well seen at higher latitudes. 
The GRS is well recovered as well as the zones (bright clouds) and belts (dark clouds), see Fig.~\ref{fig:solsys_cloud}.
The overall result is very encouraging, since we can study directly hydrodynamics of thick
atmospheres for angularly resolved Jupiter-like gas giants already now, 
with existing telescopes (see Section~\ref{sec:tel}.

\subsubsection{Neptune}\label{sec:inve_solsys_clouds_nep}

We employ the RGB Neptune map obtained with the NASA {\em Voyager} mission, 
which discovered the Great Dark Spot (GDS)
and traces of clouds at different heights in a highly windy atmosphere. 
Such spots (vortices) are long-lived but still transient features 
in the Neptune atmosphere.
As for Jupiter, we used a part of the map with the GDS at a visible latitude.
The optical geometrical albedo of Neptune is 0.41, but it is much brighter in the blue
because of Rayleigh scattering and strong methane absorption in the red wavelengths.

The GDS is well recovered but traces of clouds are below the reconstruction resolution (Fig.~\ref{fig:solsys_cloud}).
This example illustrates that blue planets may have transient vortices, which are detectable in a blue passband.

\subsubsection{Venus Clouds}\label{sec:inve_solsys_clouds_ven}

Venus is a cloud-enshrouded rocky planet, with clouds made of sulfuric acid droplets.
It is possible that quite a few rocky exoplanets at closer distances than the inner edge of the WHZ
are completely covered by clouds as Venus, with a strong green-house effect. 
Venus upper clouds have a transient structure 
driven by fast winds (jets) in its atmosphere, so the entire atmosphere rotates
in only four Earth days and in the opposite direction, as compared to the 243 Earth day surface rotation. 
We employed the RGB image of the Venus upper clouds obtained by the NASA {\em Galileo} spacecraft. 

The Venus cloud optical geometrical albedo is very high, 0.69. 
The cloud structure on Venus is different from that on Jupiter, Neptune and other gas giants (Fig.~\ref{fig:solsys_cloud}).
It is encouraging that our indirect imaging can distinguish between different cloud
structures and provide insights unto global circulation of atmospheres on rocky and
gas giant planets.

\subsection{Solid Surfaces of Planets and Moons}\label{sec:inve_solsys_surf}

\begin{figure*}
\centering
\includegraphics[width=17cm]{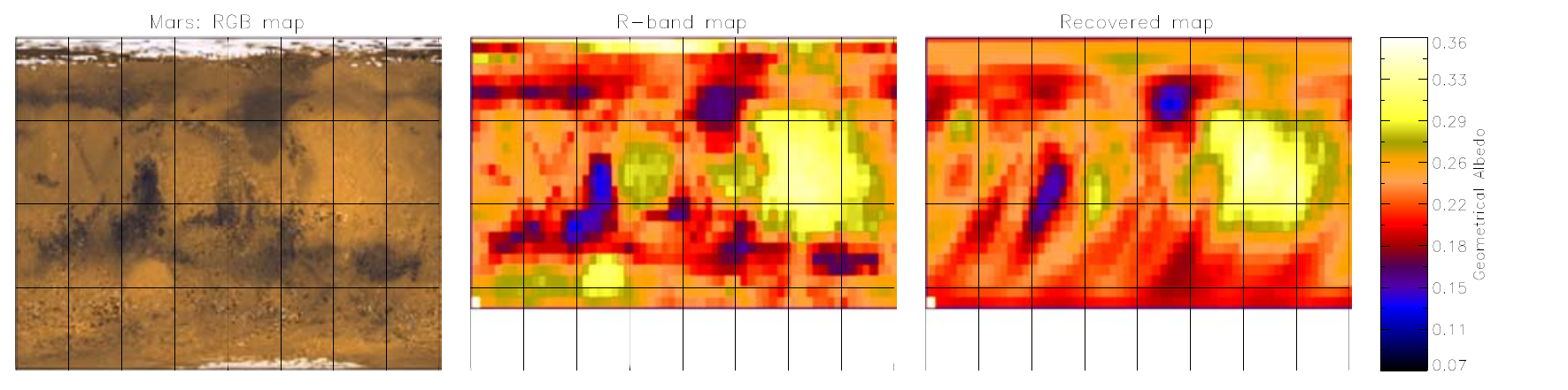}
\includegraphics[width=17cm]{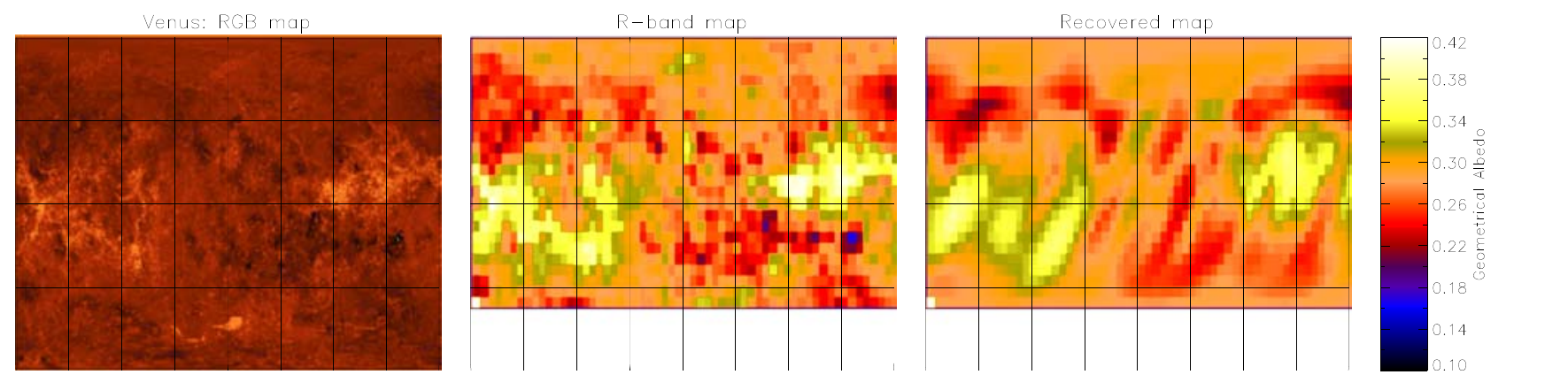}
\includegraphics[width=17cm]{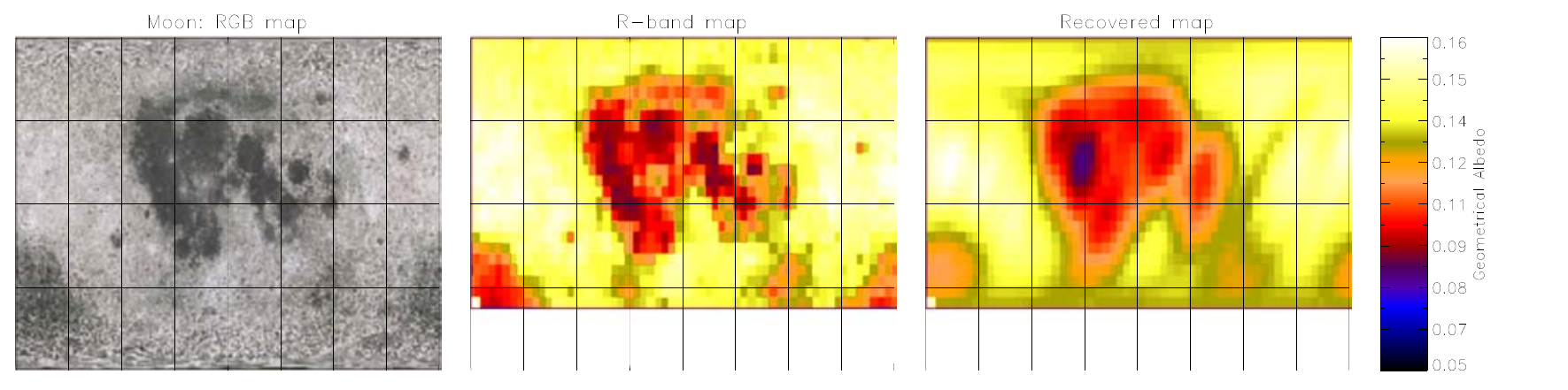}
\includegraphics[width=17cm]{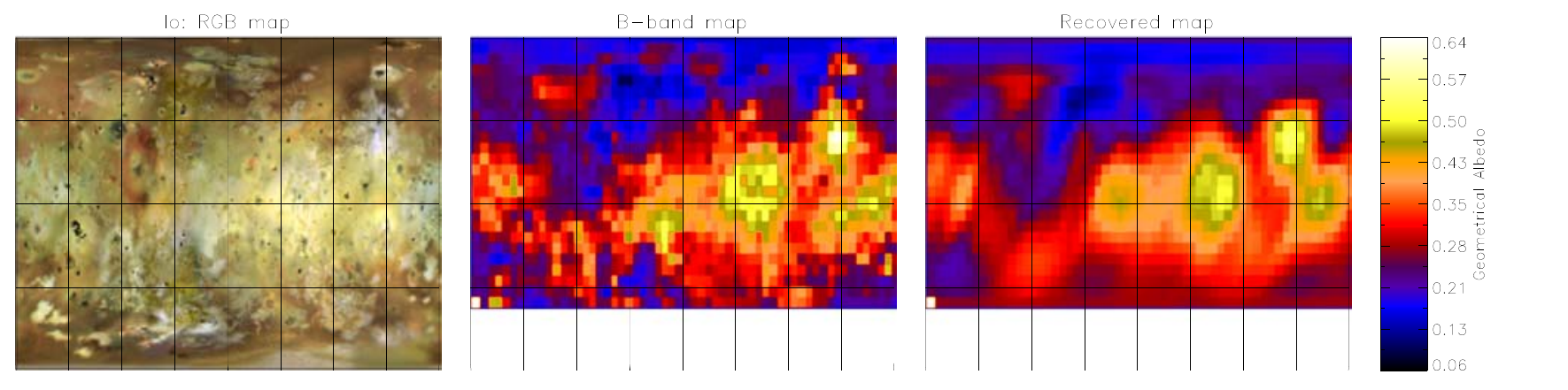}
\includegraphics[width=17cm]{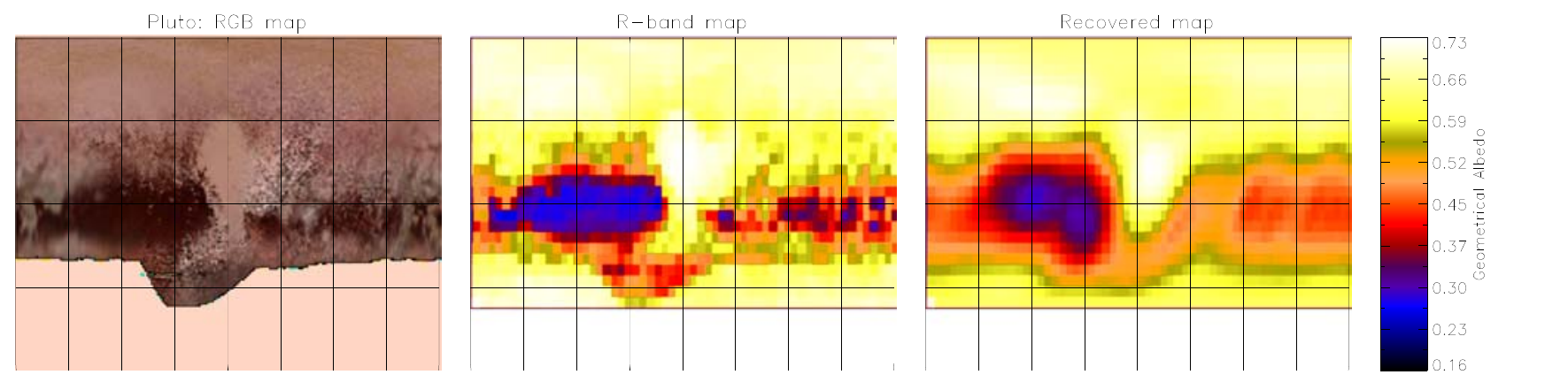}
\caption{Albedo maps for surfaces of some Solar system planets and moons. 
In the first column, are original RGB maps of higher resolution,
in the second column are original maps rebinned to the $6^\circ\times6^\circ$ grid in a particular band,
and in the third column are recovered maps in the same band. 
The assumed geometrical parameter values are  
$i_{\rm r}=60^\circ$, $i_{\rm o}=30^\circ$, and $\zeta_{\rm o}=30^\circ$.
The original rebinned maps are used to simulate the "observed" light curve from
which recovered images are obtained using inversions.
}
\label{fig:solsys_surf}
\end{figure*}

For planets with transparent atmospheres or without atmospheres at all we can
recover images of their surface albedo variations, as we demonstrated above for the Earth.
Here we look at several Solar system planets and moons with interesting surface features.
They generally reveal history of meteoroid bombardment and geological (tectonic or volcanic)
activity. Our results are compiled in Fig.~\ref{fig:solsys_surf}, and details for each
planet are presented in following subsections. The Solar system moons are used here only
to illustrate how an exoplanet with similar surface features can be studied
(detecting and imaging an exoplanet moon would generally require much larger telescopes).

\subsubsection{Mars}\label{sec:inve_solsys_surf_mars}

Mars can be considered as an analog of cold rocky planets beyond
the stellar WHZ. It has seasons, polar ice caps and a very thin, transparent 
atmosphere, allowing a clear view of the planet surface in the optical.
On the surface there are volcanoes, canyons, and traces of earlier floods.
The polar ice caps vary with seasons. The optical geometrical albedo is
quite low, 0.17.
We employed an RGB map of Mars obtained by the NASA {\em Viking} missions.
When rebinned to the $6^\circ\times6^\circ$ grid, it still shows well the largest 
volcano in the solar system, {\em Olympus Mons}, 
as well as a large equatorial canyon system, {\em Valles Marineris}.

Shapes of dark dusty areas and bright highlands, including {\em Olympus Mons}, 
are well-recovered in the R-band image. 
If the top of {\em Olympus Mons} was covered by snow, it would be quite conspicuous
(see, for example a reconstruction of volcanos on Io in Section~\ref{sec:inve_solsys_surf_io}).
Thus, seasonal variations of the albedo on exoplanets like Mars
can be studied with the EPSI.

\subsubsection{Venus Surface}\label{sec:inve_solsys_surf_ven}

The Venus surface is not visible in the optical because of thick clouds.
Its map is obtained using radar measurements. Under the clouds, the surface looks
orange because red light passes through the atmosphere. 
The surface is hot, and structures are relatively small-scale because of regular
resurfacing. Features include craters from large meteoroids, because only those can reach
and impact the surface, multiple volcanos, and some highlands.
For our reconstruction we have assumed that the surface optical albedo is the same 
as for Mercury, 0.14, and that the surface can be seen through a transparent atmosphere.

Variations of the albedo of different structures are quite well reconstructed, 
including large-scale dark and bright areas with some substructures. Volcanos are
quite small and are not resolved in the recovered image. However, if volcanic
activity affected a larger area, that could be recovered (like on the Moon or Io, see the next section). 
Thus, time series of such indirect images of active exoplanets will allow studying volcanic
activity and chemical composition of the new lava through multi-band
spectral imaging, as discussed for biosignatures and technisignatures in 
Section~\ref{sec:signat}.

\subsubsection{Moon}\label{sec:inve_solsys_surf_moon}

The Moon is an interesting analog of an exoplanet without an atmosphere.
It is highly cratered, similar to Mercury, but it also has large areas of
lighter highlands and darker maria filled with volcanic lava some billions years ago. 
The light and dark areas have different composition. The optical albedo
of the Moon is 0.12, similar to that of Mercury.

The overall shapes of the large-scale brighter and darker areas are well reconstructed.
Such variety of the albedo on exoplanets without an atmosphere may help
reveal the history of their formation and volcanic activity, as well as the composition
of various areas.

\subsubsection{Io}\label{sec:inve_solsys_surf_io}

The Jupiter moon Io can be an analog of a very volcanically active exoplanet. 
Volcanic lava fills in impact craters and spreads over
the surface, which is constantly renewed. Vivid surface colors of Io
revealed by the NASA {\em Galileo} and {\em Voyager} missions are cased by silicates and 
various sulfurous compounds. In particular, volcanic plum deposits can be white,
due to sulfur dioxide frost, or red, due to molecular sulfur. 

We have employed a detailed RGB map of Io obtained by the {\em Galileo} mission.
We made it visible from the South pole of the moon, so that interesting 
surface features are in the view. The optical geometrical albedo used is 0.45.
Large-scale, active volcanic areas are well reconstructed. In particular, 
the volcano {\em Pele} distinguished by a large circular red plum deposit 
(in the upper left part of the map) is well seen in the recovered image. 
Bright volcanic slopes covered by sulfur dioxide frost, 
such as {\em Tarsus R.} (in the upper right part of the map), are also well recovered.
We did not attempt it, but it seems possible that large volcanic plums 
forming umbrella-shaped clouds of volcanic gases can also be detected 
with our imaging technique, if they last long enough.

Discovering and imaging such geologically active exoplanets will be
valuable for studying their internal composition. Complementary images
in the infrared will help to determine temperature and age of volcanic deposits.

\subsubsection{Pluto}\label{sec:inve_solsys_surf_pluto}

Pluto is an icy, rocky dwarf planet which may serve as an analog of exoplanets
and planetesimals beyond the so-called frost line distance from the host star. 
Images of the Pluto delivered by the NASA {\em New Horizons} mission reveal spectacular 
glaciers of frozen methane, carbon monoxide, and nitrogen. 
Extended dark surface features are possibly due to tholins \citep{sag79}, which
are common on surfaces of icy outer Solar system bodies and formed by UV or energetic particle
irradiation of a mixture of nitrogen and methane. The optical geometrical albedo used is 0.49.
The surface of Pluto was found to have unexpected tectonic activity indicated by 
large craters with degradation and infill by cryomagma. 

Since the {\em New Horizons} map of Pluto is incomplete, we assigned to the missing part 
a constant intermediate value albedo. Also, we made the planet visible from the pole with
a detailed map, so that the missing part is basically invisible with a given inclination angle. 
All large-scale structures, including the brightest glacier and tholin-rich areas are well 
reconstructed. This is promising for detecting and stydying cryovolcanic activity 
on icy exoplanets.

\section{Biosignatures and Technosignatures}\label{sec:signat}

In this section, we present models for spectral (broad-band) imaging
of exoplanets, with the same parameters as in Section~\ref{sec:solsys}. 
We demonstrate that, for example, photosynthetic biosignatures and 
inorganic surface composition can be detected using EPSI. 
We also model hypothetical planets with global artificial structures 
which can be detected under certain circumstances.

\subsection{Spectral Imaging of Biosignatures}\label{sec:inve_bio}

\begin{figure}
\centering
\includegraphics[width=7cm]{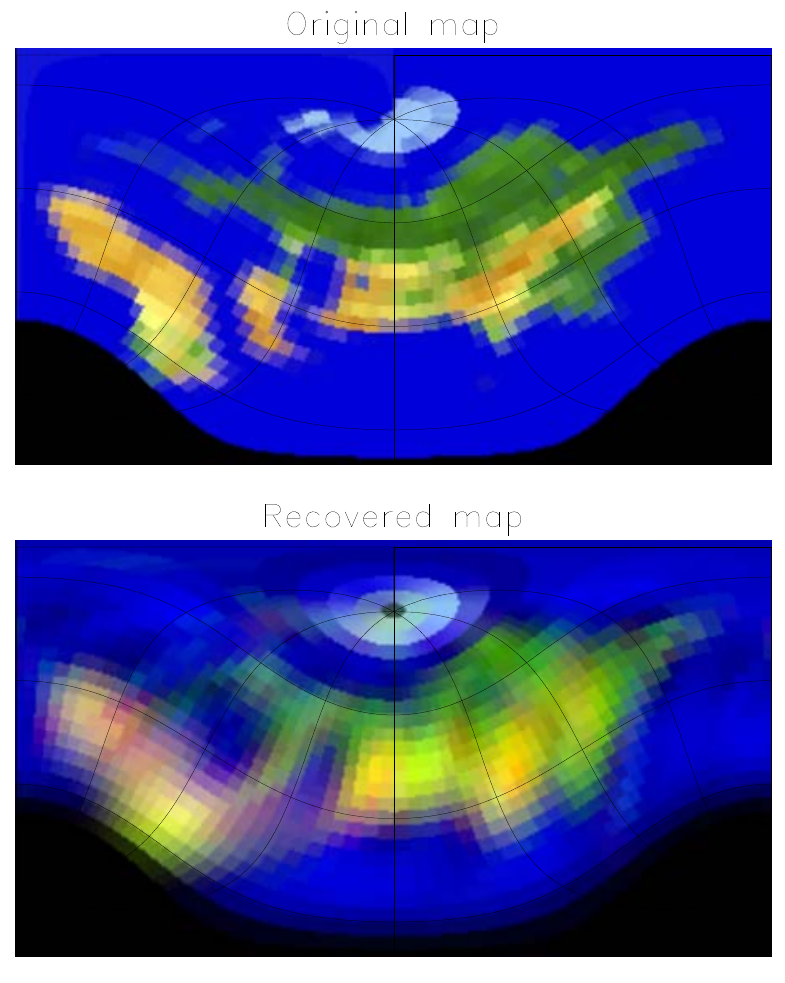}
\caption{The RGB composite original and recovered maps of an exo-Earth with continents, ocean and a polar cap.
Continents are in part are covered by green photosynthetic organisms and in part by deserts.  
}
\label{fig:EN3000_exoearth_rgb}
\end{figure}

\begin{figure}
\centering
\includegraphics[width=7cm]{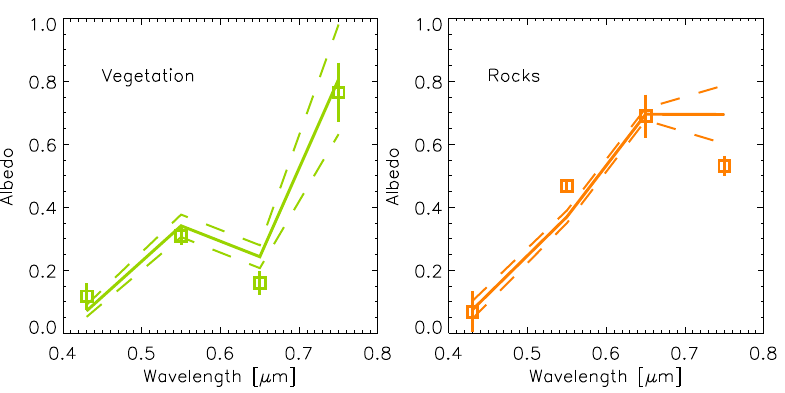}
\caption{Average spectra of selected areas from the images in Fig.\ref{fig:EN3000_exoearth_rgb}.
Left: spectra of a small green area at mid latitudes in the eastern part of the planet image.
Right: spectra of a small desert area at lower latitudes in the western part of the planet image.
Solid lines are the original image spectra. Dashed lines show a range of "natural" variations 
within the area in the original image. Symbols depict spectra from the recovered image.
The presence of chlorophyll-rich organisms is unambiguously detected. The composition of the desert surface
can be recovered by a comparison with spectra of minerals.
}
\label{fig:EN3000_exoearth_veg}
\end{figure}

\begin{figure*}
\centering
\includegraphics[width=7cm]{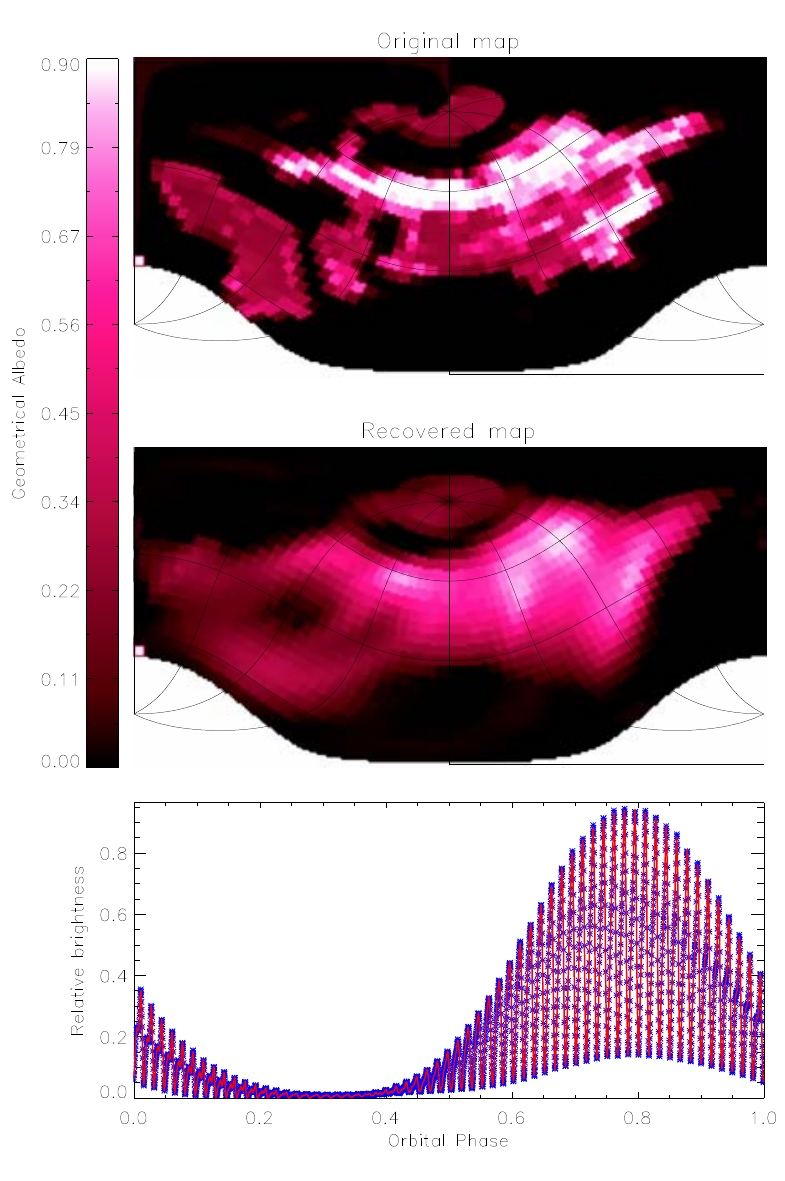}
\includegraphics[width=7cm]{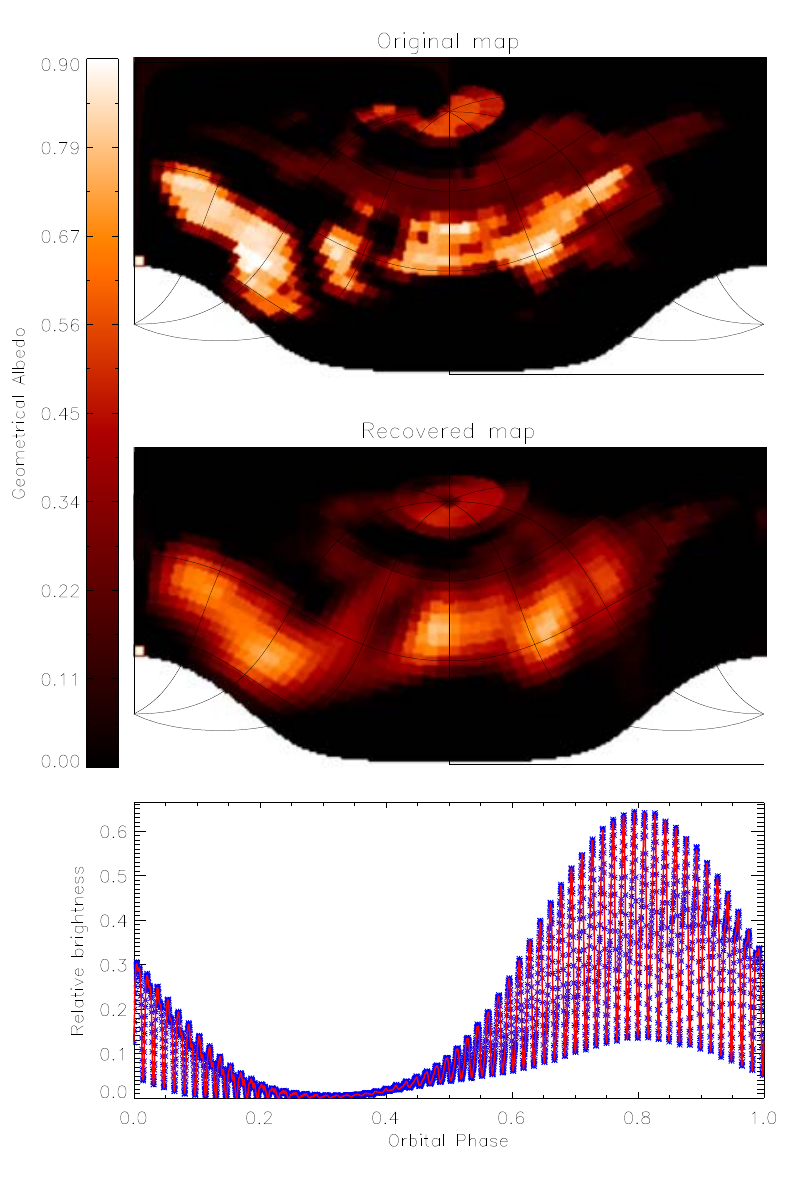}
\includegraphics[width=7cm]{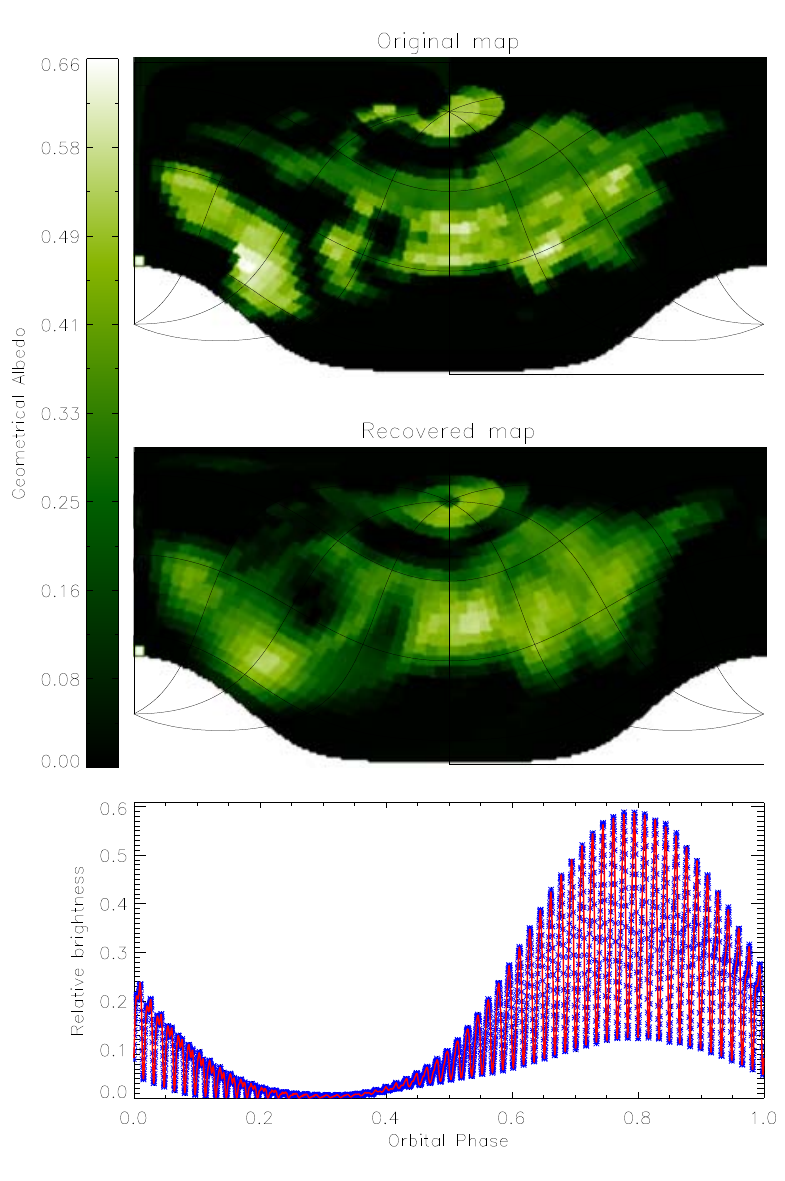}
\includegraphics[width=7cm]{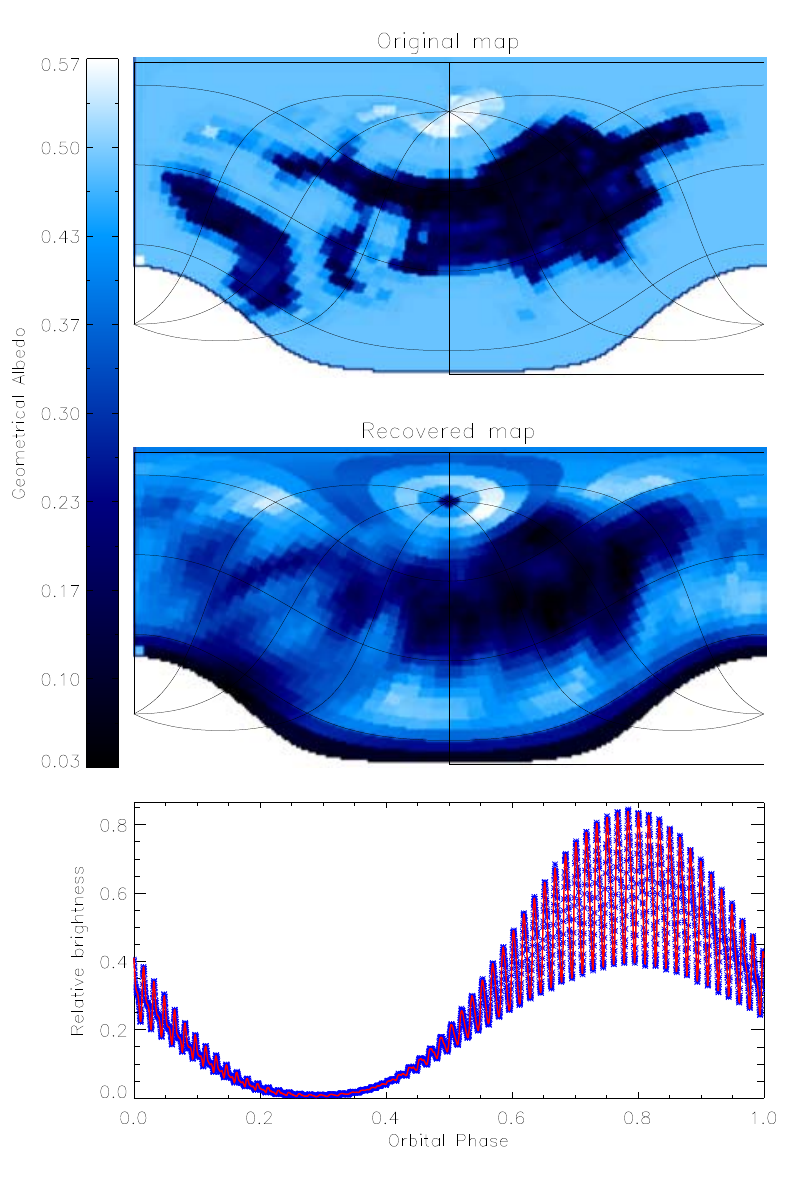}
\caption{The same as Fig.~\ref{fig:EN3000} but for the input maps in the four RGB+NIR bands from 
an exo-Earth image shown in the top panel of Fig.~\ref{fig:EN3000_exoearth_rgb}. 
The RGB composite of the recovered images is shown in the bottom panel of Fig.~\ref{fig:EN3000_exoearth_rgb}. 
The map quality indexes are 
IR: IQ=82\%, SD=16\%;
R: IQ=86\%, SD=15\%;
G: IQ=88\%, SD=15\%;
B: IQ=86\%, SD=25\%.  
}
\label{fig:EN3000_exoearth_ir+rgb}
\end{figure*}

Detecting biosignatures on exo-Earths is an ultimate goal of modern
exoplanet studies. In most cases, this requires an analysis of 
the spectral content of the reflected, emitted or/and absorbed flux
from an exoplanet. For example, detecting spectral 
signatures of biogenic (or simply out-of-equilibrium) gases in exoplanetary 
atmospheres is a promising approach \citep[e.g.,][]{gren2013}.
Positive atmospheric detections will help select promising candidates, 
but definite conclusions for exolife detection will require 
complementary studies to exclude false-positives.

Here we propose such a complementary approach using our surface imaging technique
applied to exoplanet light curves measured in several spectral bands
or using an efficient spectrograph.
We show in this Section that a "true"-color map of exoplanets
inferred with our surface imaging technique could  provide a powerful
tool for detecting surface-based living organisms,
in particular photosynthetic organisms which have dominated the Earth
for billions years.

We test our approach using a composite RGB image of Earth's surface samples:
ice polar caps, ocean reflecting blue sky, deserts, and forest. 
This image is shown in Fig.~\ref{fig:EN3000_exoearth_rgb} (top panel).
We complement this RGB image with a near-infrared (NIR) image (0.75--0.85\,$\mu$m) 
using our laboratory study of various organic and inorganic 
samples \citep{berdetal2016}. In particular, the most striking feature 
of the IR image is high reflectance by the green vegetation. 

We have carried out inversions of the four RGB+NIR images separately,
and obtained recovered images as described in this paper. The input and output
images are shown in Fig.~\ref{fig:EN3000_exoearth_ir+rgb}.
Under the favorable observing conditions, the quality of the recovered
images is quite good. When combined in a "true"-color photograph,
the recovered RGB-image outlines quite remarkably the polar ice-cap 
and continents with large desert and vegetation areas 
(see Fig.~\ref{fig:EN3000_exoearth_rgb}, lower panel).

By resolving surface features, spectra of various areas can be 
extracted and explored for life signatures. 
An example average spectrum
of a small green area at mid-latitudes in the eastern part of the planet
is shown in the left panel of Fig.~\ref{fig:EN3000_exoearth_veg},
for both the input and recovered images. A geographically resolved spectrum  
with a deep absorption in the visible and high albedo in the NIR 
has been recovered very well. 
It reveals quite strikingly the presence of the
chlorophyll-rich organisms with the distinct red-edge signature typical for
green plants and cyanobacteria. Thus, using spatially resolved images 
of exoplanets dramatically increases our chances to detect exolife, because
we can carry out a spectral analysis of areas with high concentrations of
living organisms, which is necessary for 
unambiguous detection 
\citep[see models with various biopigments in][]{berdetal2016}. 

In the right panel of Fig.~\ref{fig:EN3000_exoearth_veg}, we show 
an example spectrum of a desert area, whose composition can be
recovered using spectra of various minerals.

\subsection{Artificial Mega-Structures of Advanced Civilizations}\label{sec:inve_art}

\begin{figure}
\centering
\includegraphics[width=7cm]{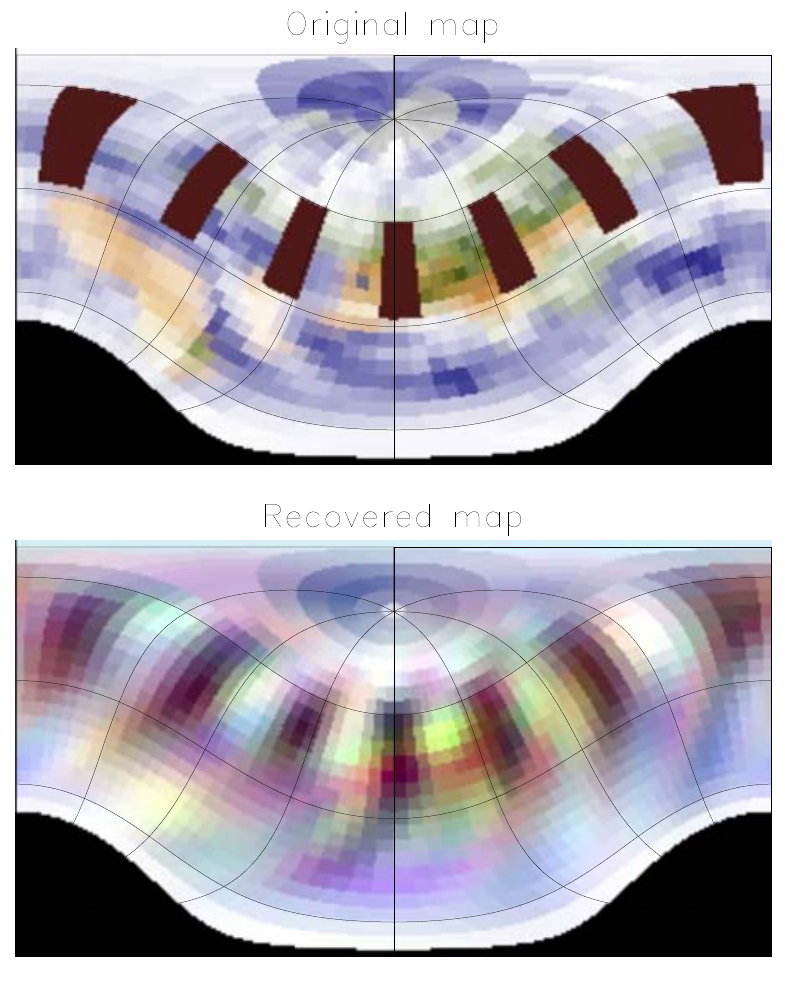}
\caption{The RGB composite original and recovered maps of an exo-Earth with the 
artificial mega-structure of low albedo (0.05) above clouds, imitating a 
photovoltaic-like power-plant in space.
}
\label{fig:EN3000_exoearth_ams_black}
\end{figure}

\begin{figure}
\centering
\includegraphics[width=7cm]{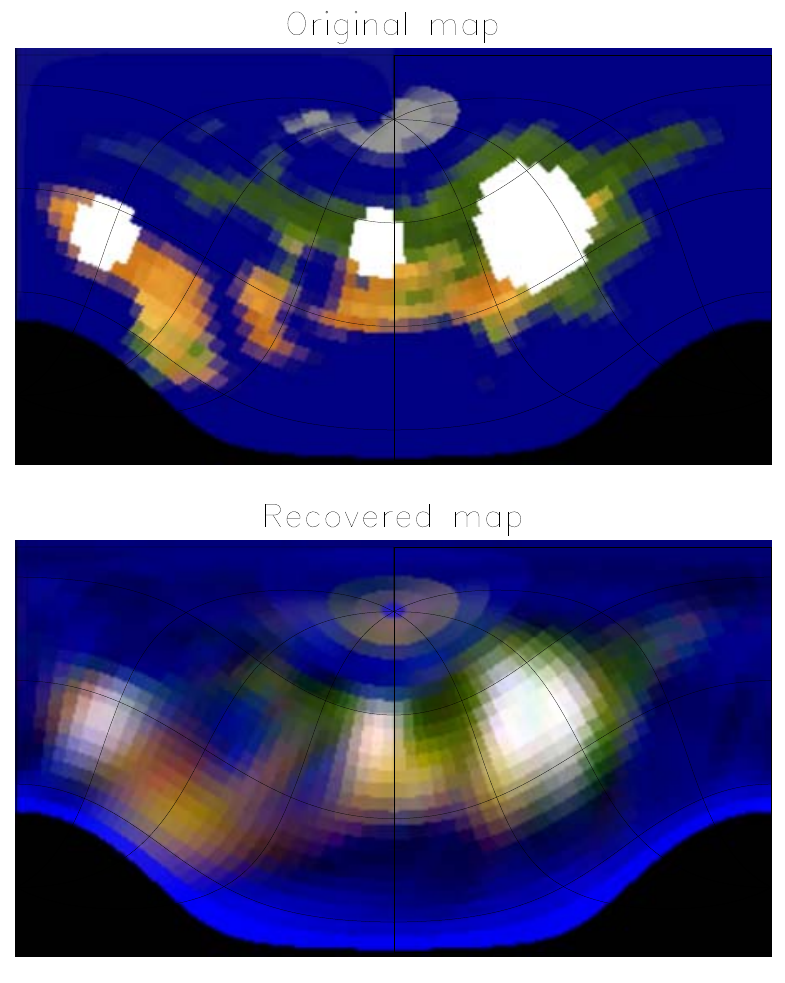}
\caption{The RGB composite original and recovered maps of an exo-Earth with the 
artificial mega-structure of high albedo (0.9), imitating urban-like areas under 
reflective "umbrellas" (white circles).
}
\label{fig:EN3000_exoearth_ams_white}
\end{figure}

\begin{figure}
\centering
\includegraphics[width=7cm]{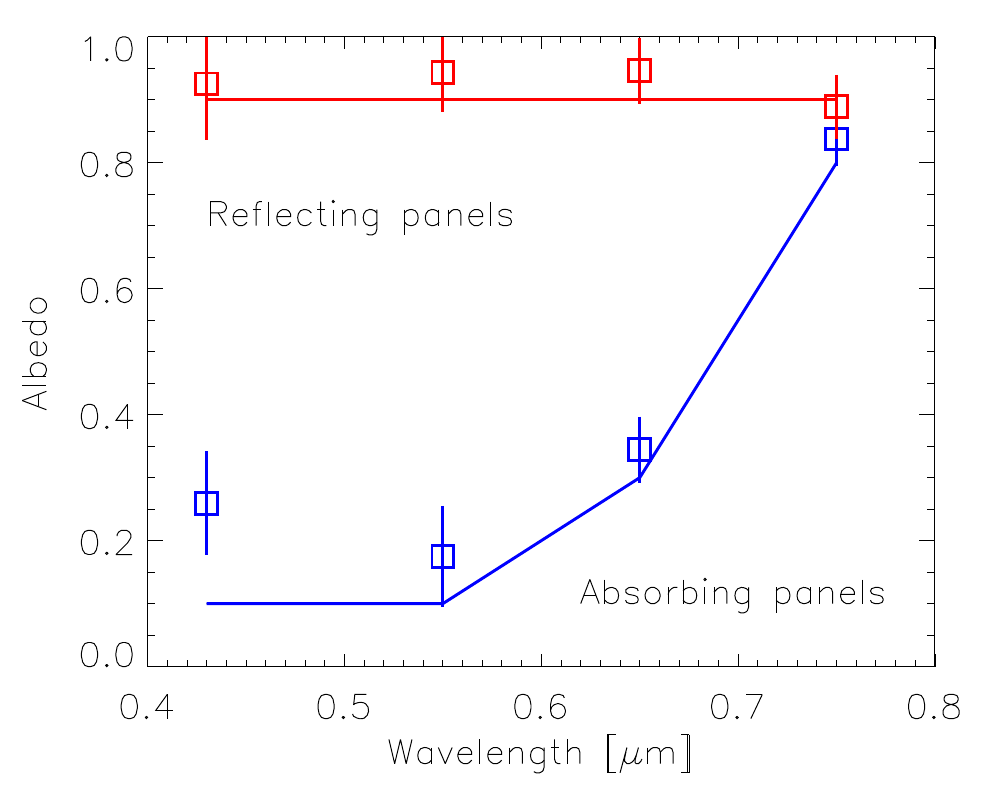}
\caption{Average spectra of the artificial mega-structures 
shown in Figs.~\ref{fig:EN3000_exoearth_ams_black} and ~\ref{fig:EN3000_exoearth_ams_white}.
The solid lines and symbols show original and recovered spectra, respectively.
Red and blue colors stand for spectra of the high- and low-albedo, respectively. 
}
\label{fig:EN3000_exoearth_ams_sp}
\end{figure}

The EPSI technique allows also for detecting artificial mega-structures (AMS)
constructed by advanced civilizations
either on the surface or in the near-space of an exoplanet. 
We consider AMS to be of some "recognizable" shape and/or 
homogeneous albedo, and most probably to be "geostationary". We have previously
described how the thermal footprint of a civilization only slightly more advanced than that of Earth's
could be detectable at IR wavelengths from its planetary rotation signature \citep{kuhnberd2015}. Since
advanced civilizations would depend on stellar power sources to avoid planetary warming, it is likely that the
low albedo stellar energy collectors should be detectable using the mapping technique described here.

One example of such low-albedo installations is similar to our photovoltaic systems 
on the Earth's surface and in space. 
Such structures would absorb stellar light with high efficiency 
in particular spectral bands.
In fact, spectral signatures of such alien photovoltaics 
could be of similar shapes to those of various biopigments 
(see Fig.~\ref{fig:EN3000_exoearth_veg} and \cite{berdetal2016}), 
depending on the alien technology, energy needs, and the available stellar light 
before or after it passes through the planetary atmosphere.

Another example would be high-albedo installations, also on the surface or in the near-space,
in order to redirect the incident stellar light, e.g., for heat mitigation 
by reflecting the light back to space.
Such installations may reflect only a particular part of the spectrum, 
because the amount of the reflected energy can be regulated by both the size 
of the structure and the width of the spectral band.

We have simulated examples of exoplanets with AMS of the high and low albedo
combined with natural environment, similar to the Earth. Light curves were simulated
in the RGB passbands and inverted separately, as described in previous sections 
for natural albedo variations. The final images were obtained
by combining the individual RGB light curve images into "true"-color images.

Our model AMS are shown in 
Figs.~\ref{fig:EN3000_exoearth_ams_black} and ~\ref{fig:EN3000_exoearth_ams_white},
imitating a photovoltaic-like power-plant in space, above clouds, 
and urban-like areas under reflective "umbrellas" on a planet surface 
under a cloudless atmosphere.
These numerical simulations show that recognizing AMS in inferred images 
requires quite large areas covered by AMS, because of uneven illumination 
and limited spatial resolution. 
Also, higher contrast structures with respect to the natural environment 
are obviously much more conspicuous and easier to infer than structures with the albedo
similar to that of the environment. 

Distinguishing such AMS from the high and low albedo natural environments 
(e.g., bright ice caps and mountain tops, dark lava fields and water reservoirs, etc.) 
may be a challenge unless they have a regular structure (like a power-plant).
Also, analyzing the spectral content of the reflected light can be useful. 
As mentioned above, the reflectance of AMS can be limited to particular wavelengths.
As examples, in Fig.~\ref{fig:EN3000_exoearth_ams_sp} we show the input and recovered 
spectra of the AMS with absorbing and reflecting panels. The stationary nature of AMS
could also be an important factor in their identification.

\section{Proxima b and the Alpha Centauri system}\label{sec:acen}

The Alpha Centauri system stars A and B and Proxima Centauri are the closest stars
to the Sun. There is at least one exoplanet discovered so far
in this system -- Proxima b \citep{proxb}. Gas giant, Jupiter-like planets are
concluded to be incompatible with the present measurements \citep{acen_jup}.
Proxima b is an excellent candidate for first-time
exoplanet surface imaging. 

\begin{figure}
\centering
\includegraphics[width=7cm]{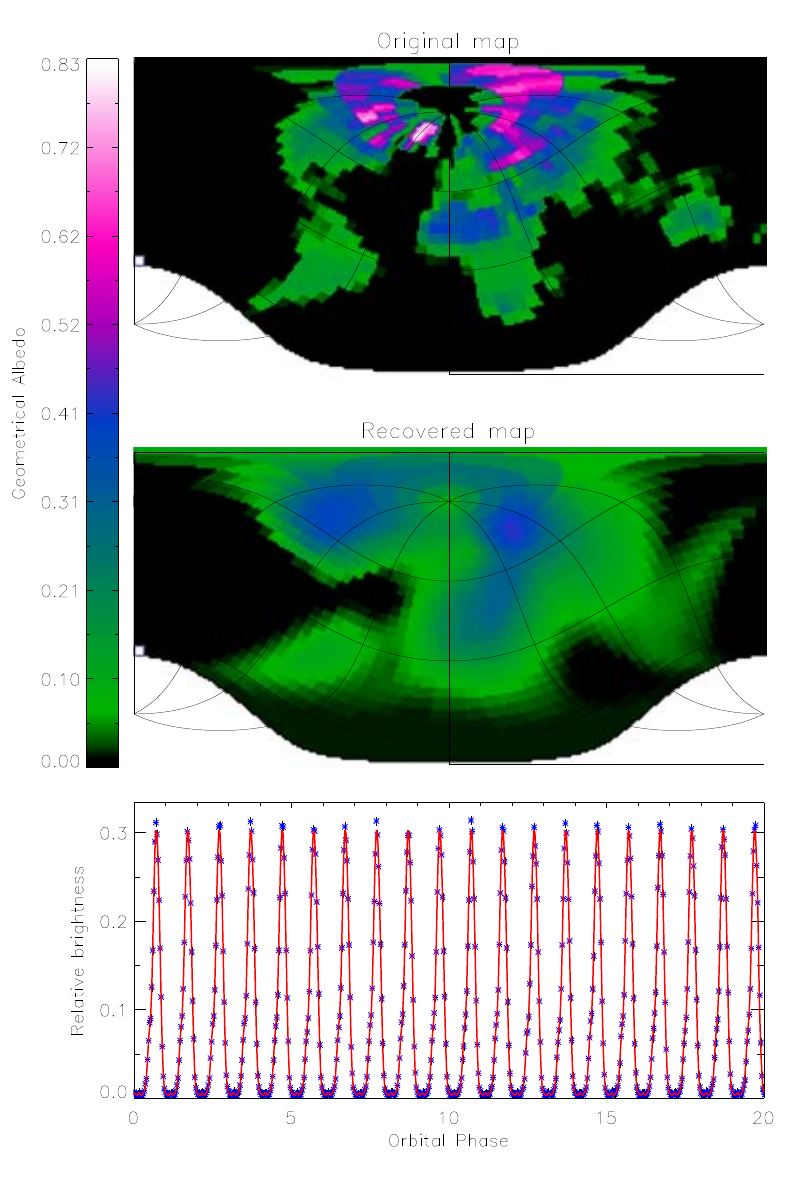}
\caption{The same as in Fig.~\ref{fig:EN3000} for the Proxima b model with S/N=100
and the planet rotation being locked to the orbital motion at the 1:1 resonance.
The light curves is simulated as it would be observed during several (20) orbital periods
with 30 phase measurements per period, i.e., $N$=600.
Here IQ=80\%\ and SD=13\%.}
\label{fig:prox11}
\end{figure}

\begin{figure}
\centering
\includegraphics[width=7cm]{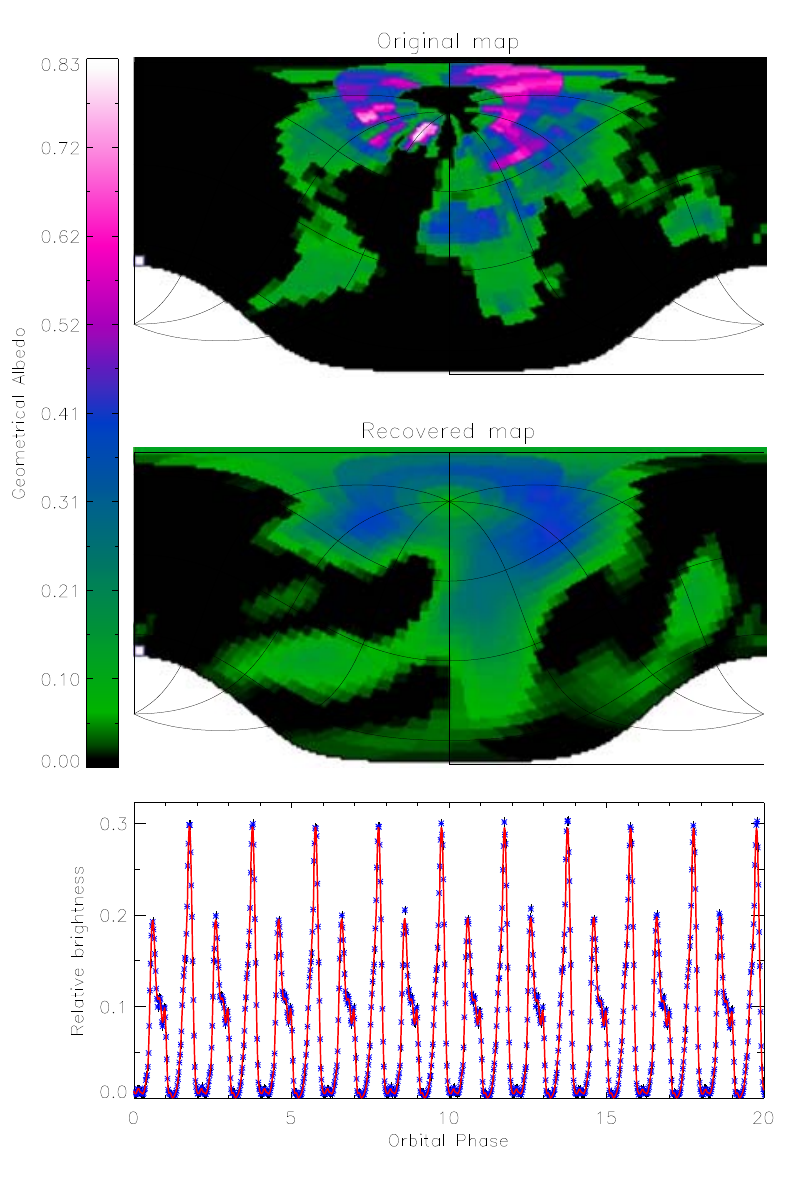}
\caption{The same as in Fig.~\ref{fig:prox11} for the 3:2 resonance.
Here IQ=80\%\ and SD=13\%.}
\label{fig:prox32}
\end{figure}

Proxima b orbits the M5 red dwarf within its WHZ
with the period of about 11.2 days at about 0.04 AU from the star. 
At this close distance, tidal interactions may have locked the axial
rotation of the planet to its orbital period or to the 3:2 resonance as 
for Mercury. The presence of a dense atmosphere may
influence the tidal interaction too \citep{prox_atm}. 

In this section, we present simulations and inversions for the Proxima b
planet assuming tidally locked orbits at the 1:1 and 2:3 resonances.
with the planetary axis inclined with respect to the orbit plane.

Since Proxima b was discovered using the radial velocity (RV) technique
and so far no transits were detected, the inclination of the orbit is still
unknown. Thus, the planet can be 
an Earth-like planet of $1.3-1.5 M_{\rm E}$ at $i_{\rm o}$ of $90^\circ-60^\circ$, 
a super-Earth of $1.6-5 M_{\rm E}$ at $i_{\rm o}$ of $55^\circ-15^\circ$,
or a Uranus/Neptune-mass planet of $15-17 M_{\rm E}$ at $i_{\rm o}\approx5^\circ$.
Extremely low orbit inclinations $i_{\rm o}<5^\circ$
(a nearly face-on orbit) are still possible with a few percent chance, which
may lead up to a Saturn-mass planet at $i_{\rm o}\approx1^\circ$. 
The composition and, hence, the surface (or cloud) albedo of such planets may differ 
significantly. Therefore, obtaining albedo maps (especially at different wavelengths/passbands)
can provide a powerful constraint on the planet mass and composition as well as
its potential habitability. 

As in Section~\ref{sec:inve_noclouds}, we test our inversion technique for Proxima b parameters
using cloudless Earth's albedo maps. We assume the orbit inclination
of $60^\circ$ corresponding to a 1.5\,$M_{\rm E}$ planet. With the terrestrial mean density 
of 5.5\,g/cm$^3$ its radius would be 1.15\,$R_{\rm E}$, which we use in our reflected light 
simulations. Here, the SNR=100 is assumed. 

The two scenarios when the planet rotation is 
locked to the orbital motion at the 1:1 and 3:2 resonances are shown in Fig.~\ref{fig:prox11} 
Fig.~\ref{fig:prox32}, respectively.
The fact that the periods are the same or very close hinders the amount of information
in the light curve, even though we allow observations during several orbital periods.
However, most of the landmass (80\%) is still recovered and its outline
is quite recognizable. The quality of the map is higher for the 3:2 resonance orbit
because there are two different light-curves observed every other orbital period.
Here, the number of measurements and the SNR can be increased by  
observing during several orbital periods, as shown in our inversions.

\begin{figure}
\centering
\includegraphics[width=7cm]{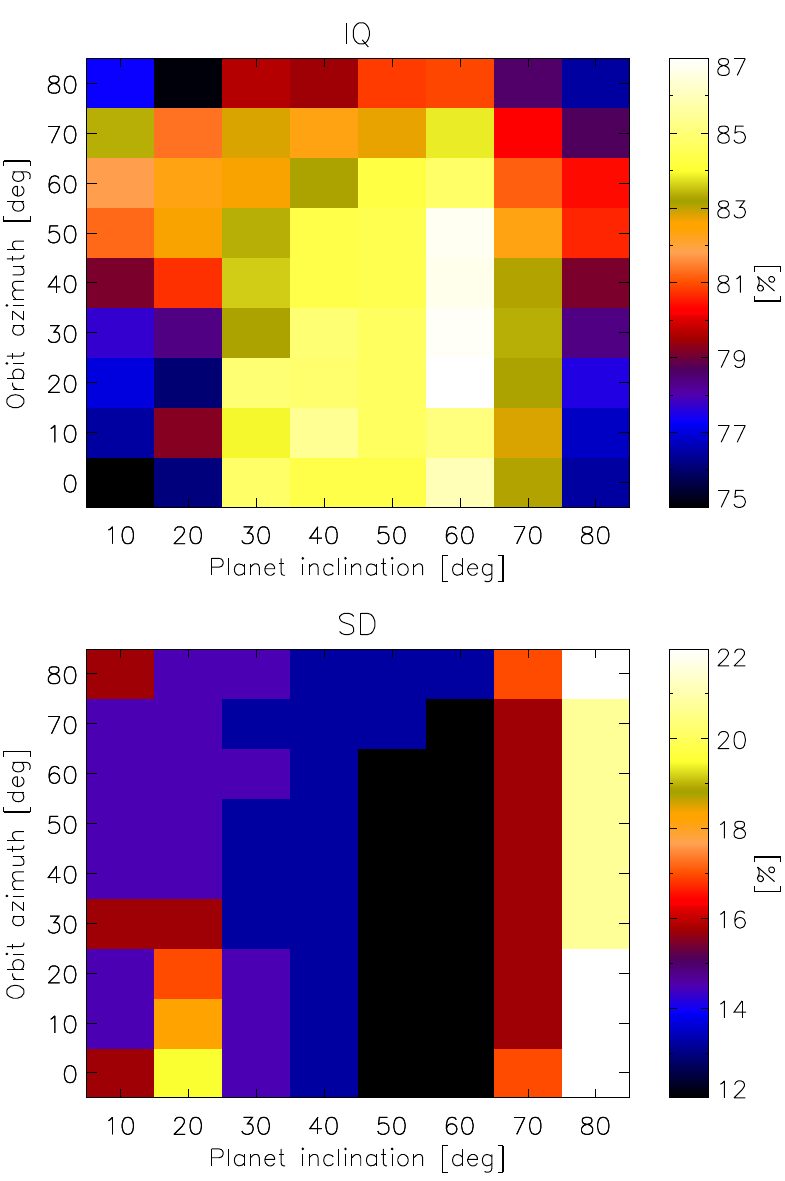}
\caption{Dependence of the IQ parameter (Eq.~\ref{eq:iq}) and map standard deviation SD
on the planet axis inclination angles $i_{\rm r}$ and the orbit normal azimuth 
$\zeta_{\rm o}$ for the Proxima b model with the periods locked at 1:1. 
High values are in light tones, and low values are in dark tones.
The median values of IQ and SD are 80\%\ and 16\%.}
\label{fig:prox_iq}
\end{figure}

Since the rotational axis inclination angle and the orbit normal azimuth
are unknown, we vary them to understand imaging capabilities of this planet,
possibly locked at a resonance rotation.
The IQ and SD parameters of our inversions for the periods locked at 1:1 are shown in
Figs.~\ref{fig:prox_iq}. Similar to the unlocked case, the most favorable planet axis inclination
angles range within $30^\circ$ to $60^\circ$, while the orbit normal azimuth angle
can be in a wide range.

\section{Observational Requirements}\label{sec:tel}

In this section, we investigate and formulate observational requirements
for the telescope size and scattered light level, in order to obtain spectral images
of Proxima b and possible rocky planets in the nearest to the Sun stellar system 
Alpha Centauri A and B.  
We also estimate how many Earth-size and super-Earth planets in WHZ can be imaged with our 
EPSI technique depending on the telescope aperture for a given scattered light 
background.

\subsection{Contrast Profile}\label{sec:contrast}

\begin{figure}
\centering
\resizebox{7cm}{!}{
   \includegraphics{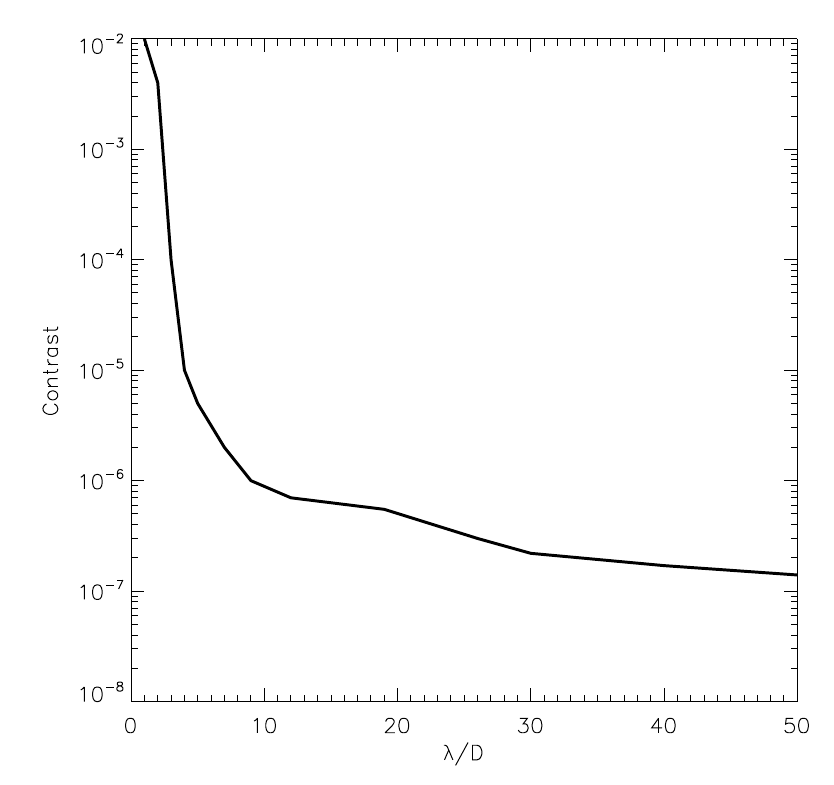}
   }
\caption{An assumed contrast curve for the scattered stellar light brightness 
versus the angular distance from the star in units of the ratio 
of the wavelength $\lambda$ to the telescope diameter $D$ 
(for an unsegmented mirror telescope).
The contrast reaches 10$^{-7}$ at $\lambda/D=100$.
}
\label{fig:psf}
\end{figure}

The level of the stellar scattered light, or contrast $C$
with respect to the central star brightness, 
plays an important role in the planet reflected light SNR. 
The required SNR has to be achieved above the bright background at 
the WHZ angular distance from the star, where an Earth-like planet may reside. 
This component of the SNR budget is missing in the computations 
made by \cite{fk12} and leads to unrealistic sensitivities.

For reference, we take the contrast curve similar to 
the existing state-of-the-art performance achieved with
the SPHERE imaging instrument at VLT \citep{sphere14,sphere17} 
using adaptive optics (AO), and spectral and angular differential imaging. 
We have also attempted to semi-quantitatively 
include the effects of wavefront phase errors, i.e., "speckle noise" \citep{AimeSoummer2004} 
by scaling this coronagraph/AO-limited curve to other telescopes.
A representative contrast profile that varies with the angle $\theta$ from 
the central star, expressed in units of the ratio of the wavelength 
$\lambda$ to the diameter $D$,  is shown in Fig.~\ref{fig:psf}. 

We emphasize that this contrast curve is achieved for a telescope with an unsegmented primary mirror. 
For a highly segmented primary mirror, without heroic segmented coronagraphy and speckle nulling AO, 
the contrast curve suffers from scattering off the segment edges and may not reach even 10$^{-5}$ contrast
at the required angular separation. 

We expect that the contrast of Keck-style telescopes 
(with a multi-segmented primary mirror and a single secondary mirror)
will be worse than that of a unsegmented telescope by approximately a factor equal to 
the number of mirror segments in the full aperture. For example, we scale up 
the SPHERE $C$ profile with the factors of 50, 500, and 800 
to obtain rough estimates of SNR and number of detectable WHZ planets 
using a Keck-, TMT- and EELT-like telescopes with segmented primary mirrors.

On the other hand, recently proposed hybrid interferometric telescopes, 
such as the 74m Colossus or 20m ExoLife Finder (ELF), 
can achieve contrast performance comparable to or even better than that of single-mirror telescopes 
\citep{kuhn2014,moretto2014}, at least in selected parts of their field of view (FOV).
Such an optical system consists of an array of off-axis telescopes from a common
parent parabola. Each off-axis telescope consists of the active primary and adaptive secondary mirrors.
They build a common focus pupil image with the angular resolution equivalent to the diameter 
of the telescope array. 
Analogous to the nulling interferometry, independent off-axis apertures can be phased 
to "synthesize" a PSF with a dark hole of extremely low scattered light and speckle noise. 
This dark hole can be moved within the FOV to follow an exoplanet orbiting the star 
by readjusting the aperture phases.
Therefore, for simplicity, we assume here that the contrast curve shown in Fig.~\ref{fig:psf} 
is also valid for a Colossus/ELF-type telescope at the location of an exoplanet. 

Another improvement of the telescope/coronagraph performance can be achieved by
using polarimetry to enhance the contrast of polarized light reflected from a planet above the stellar
background. This technique has been employed for detection and detailed studies of 
circumstellar disks \citep[e.g.,][]{kuhn01,opp08,poldisks_sphere,poldisks_gpi} 
as well as exoplanets unresolved from their stars \citep{berdetal2008,berdetal2011}. 
The direct imaging instruments SPHERE at VLT and GPI at Gemini have this option 
and are successful in imaging circumstellar disks and resolved companions.

An important contribution to the SNR budget for ground-based telescopes 
is the brightness of the sky. We have used sky magnitudes for the dark time at 
the ESO Paranal Observatory, Chile, by \cite{pat08}, while averaging variations due to the
11-year solar cycle. 
We note that the sky contribution is more important 
for fainter stars observed with smaller telescopes. As the aperture of the telescope
increases, the size of the resolution element decreases, and the sky contribution 
decreases as well.

Thus, we compute the number of photons from three different sources: the planet itself, 
the scattered stellar background at a given angular separation from the star, 
and a sky background area equal to the resolution element of a given telescope aperture. 
In this way, in the following sections, we compute SNR for exoplanets 
in the standard Johnson UBVRI bands depending on the size of the telescope,
using known stellar and planetary parameters.

\subsection{SNR for Proxima b}\label{sec:snr}

\begin{figure}
\centering
\resizebox{8cm}{!}{
   \includegraphics{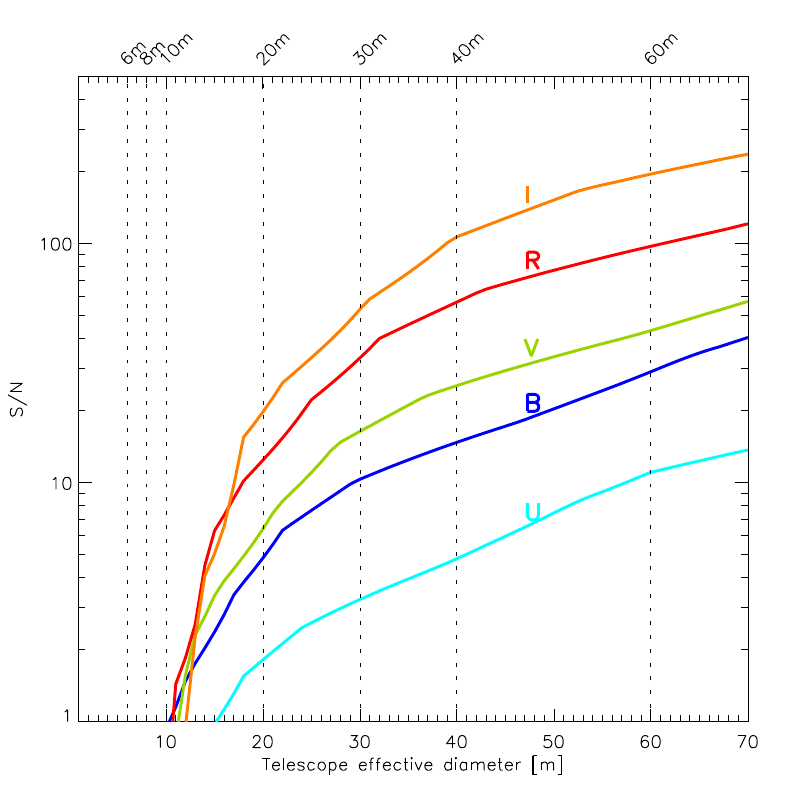}
   }
\caption{
SNR curves for an Earth-like planet like Proxima b 
(see parameters in the text)
depending on the effective telescope diameter, when assuming a single, 
unsegmented primary mirror. The SNR is computed for 1h exposure time in 
the UBVRI passbands, when only a half of the planet is illuminated.
The assumed contrast curve for the stellar scattered light 
is shown in Fig.~\ref{fig:psf}. 
The planet average surface albedo in all bands is assumed to be 0.2 for simplicity,
which allows to scale the SNR for other albedo values.  
Vertical dashed lines indicate effective diameters (unsegmented) of the existing
and planned telescopes. The resolving power of the telescopes is assumed defined
by the effective diameter.
}
\label{fig:prox_snr}
\end{figure}

Achieving SNR=100 in the reflected light from an exoplanet 
is rather challenging even for the nearest exoplanet,
such as Proxima b. Here we compute SNR in the UBVRI passbands 
for 1\,hr exposure time, when assuming the telescope efficiency 25\%, 
the illumination phase of the planet 0.5 (maximum elongation from the star),
and the average surface albedo in all bands 0.2. The stellar magnitudes
used are from the SIMBAD database. The result is shown in Fig.~\ref{fig:prox_snr}. 

Such a SNR computation produces photon-limited estimates which can be considered 
as some realistic upper limit values. Systematic errors are usually specific to 
a particular instrument design and calibration and data reduction procedures. 
Hence, we assume here the best case scenario, when all other errors are reduced 
down to the photon-noise level. In practice, to detect
reflected light variability, one should also allow for smaller illumination phases 
(down to at least 0.1) and smaller surface albedo (down to 0.05). For example,
the reflectance of vegetation in the optical is about 0.05.
The reflectance of rocky terrains and deserts is between 0.1 and 0.15, and that
of the ocean is 0.1 in the blue (due to sky reflection) and less than 2\%\ 
in the red and NIR. The glint on the ocean due to Fresnel sunlight reflection
is very bright, but its relative contribution with respect to landmasses
is low. 

We also note that because of the orbit inclination to the line of sight, 
the star-planet projected angular separation varies with the orbital phase,
leading to different $C$ levels at different orbital phases. 
Hence, orbital phases near conjunctions will be hindered by higher $C$ 
and, therefore, lower SNR for the same exposure time. 
For example, for the assumed orbit inclination of 60$^\circ$ for Proxima b
the angular separation between the star and the planet varies
by a factor of two, i.e., 
between 37\,mas at maximum elongations and 18\,mas near conjunctions.
Thus, observing Proxima b near conjunction phases may require a telescope
system with a larger resolving power or a better contrast at smaller
$\theta$ angles, depending on the orbit inclination.

The overall conclusion from this exercise is that a telescope 
with low scattered light, the effective diameter $D\ge12$\,m, 
and an equivalent or larger angular resolving power is required 
for detecting an Earth-like Proxima b above the star and sky 
background in the BVRI bands with SNR$\ge$2 during 1,hr exposure time.
A large telescope ($\ge20$\,m) is preferable
for more efficient observations.

The BVRI band measurements are particularly interesting for detecting possible 
photosynthetic life signatures as described in Section~\ref{sec:inve_bio} 
and \cite{berdetal2016}. A 20m-class telescopes
should be able to achieve SNR 5 to 20 in the BVRI-bands in 1h, at least near
maximum elongations of the planet. The R-band measurements during several 
orbital periods with SNR=20 will be already usable for surface inversions. 
With such a dedicated telescope in Chile, Proxima b can be observed between 
3 to 7 hours/night during 6 months (February to July). 
Even when discarding 10\%\ observing time for bad weather condition, 
one can obtain about 800 1h-measurements spread over 16 orbital periods 
within just one season. Observing in at least three adjacent passbands 
simultaneously (e.g., BVR or VRI) should be possible for the successful 
operation of the adaptive optics system. 

Hence, after only one season of observations we will be able to obtain
the first "color photographs" of Proxima b. If the planet is partially cloudy,
several observing seasons are needed to filter out the cloud noise
and obtain more detailed surface maps. If planet is completely covered
by thick clouds or its surface is completely featureless, we will be able
to conclude on its bulk properties already after a few orbital periods
(1--2 months).

\subsection{SNR for Alpha Centauri A and B potential planets}\label{sec:snr_ab}

A search for planets around the Alpha Centauri A and B components
has already ruled out Jupiter-like planets in their WHZ \citep{acen_jup}, but there is
still hope for smaller planets, which were not yet detected or ruled out 
due to sensitivity limits. If any planets in WHZ of these stars will be found,
they will be excellent targets for our indirect surface imaging.

\begin{figure}
\centering
   \includegraphics[width=8cm]{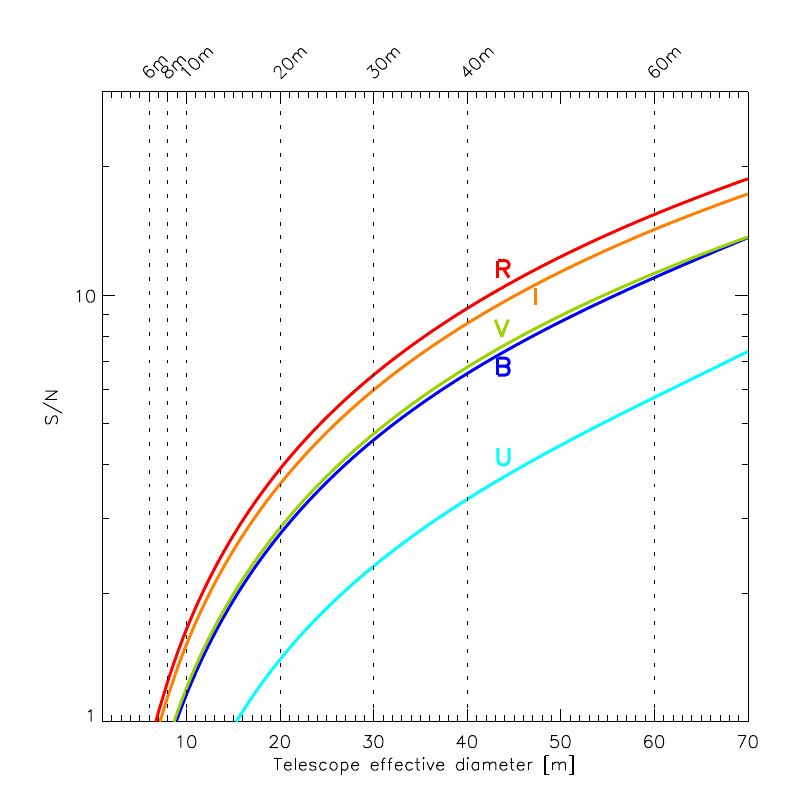}
   \includegraphics[width=8cm]{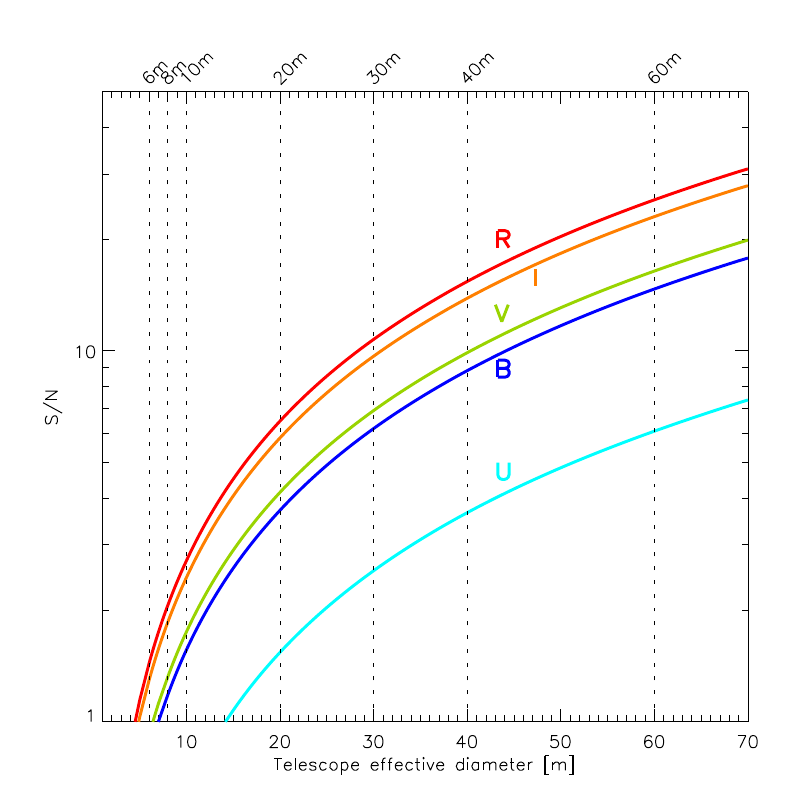}
\caption{The same as Fig.~\ref{fig:prox_snr} but for $\alpha$ Cen A and B
(top and bottom) hypothetical Earth-size planets residing 
in their corresponding (liquid water) habitable zones.
}
\label{fig:acen_ab_snr}
\end{figure}

We have computed the SNR diagram for an Earth-size planet with an average
geometrical albedo 0.2 and at the half illumination orbital phase within the WHZ 
of the Alpha Centauri A and B stars -- about 930\,mas and 560\,mas, respectively
(see Fig.~\ref{fig:acen_ab_snr}).
Resolving the WHZ of these stars is theoretically possible
even with a 1m class telescope, but achieving a significantly higher contrast 
level is needed for detecting Earth-size planets 
around these more luminous stars.
Hence, such planets may remain undetectable in the optical or NIR 
with 8--10m class telescopes, but they can be easily detected with a 20m-class telescope, 
assuming the contrast curve shown in Fig.~\ref{fig:psf}.

Our result suggests that formally, with a longer exposure time ($\ge$10h), 
an Earth-like WHZ planet may be detected around Alpha Cen B with an 8\,m telescope 
at SNR=5. In practice, however, systematic errors accumulated over several nights may prevent 
achieving the photon-limited SNR. Also, if the planetary orbit is coplanar with the binary (AB) orbit,
i.e., $i_{\rm o}\approx80^\circ$, such a detection is only possible at planet maximum elongations
from the star. Near the "full-moon" phase the planet-star angular separation is reduced by a factor 
of almost 6, which leads to a severe increase of the stellar light background and makes such
a planet undetectable. Considering that the orbit orientation on the sky plane is also
unknown, the direct detection of an Earth-size planet in the WHZ of Alpha Cen B with an
8m-telescope may take a long time.

Similar to the Proxima b case discussed in the previous section, 
dedicated continuous direct imaging observations of the Alpha Centauri A and B stars
with a larger telescope
may reveal (or exclude) Earth-size planets in their WHZ. 
Obviously, larger planets can be detected within shorter observing time,
as the SNR increases with the size of the planet.
The advantage here is that these planets will most probably have axial rotation periods 
different from the orbital periods (not tidally locked), 
which will allow for a more detailed surface mapping as presented in Section~\ref{sec:sim}.

\subsection{Number of Detectable Exoplanets}\label{sec:npl}

Assuming the telescope and instrument performance as in Section~\ref{sec:snr}, 
we can evaluate how many
WHZ Earth-size planets in the solar neighbourhood can be potentially detected
and surface-imaged. From the SIMBAD database, we have compiled a list of AFGKM
main sequence stars brighter than $V=13$ magnitude (about 3500 stars). 
We have computed their WHZ distances by scaling the 
Sun-Earth WHZ with the stellar luminosity for an Earth-like planet.
The reflected light flux is computed using the Earth size, 
geometrical albedo 0.2 for all passbands, and 0.5 illumination phase.
We consider a planet detectable if SNR$\ge$5 can be achieved during 4h exposure time.

\begin{figure}
\centering
\resizebox{7cm}{!}{
   \includegraphics{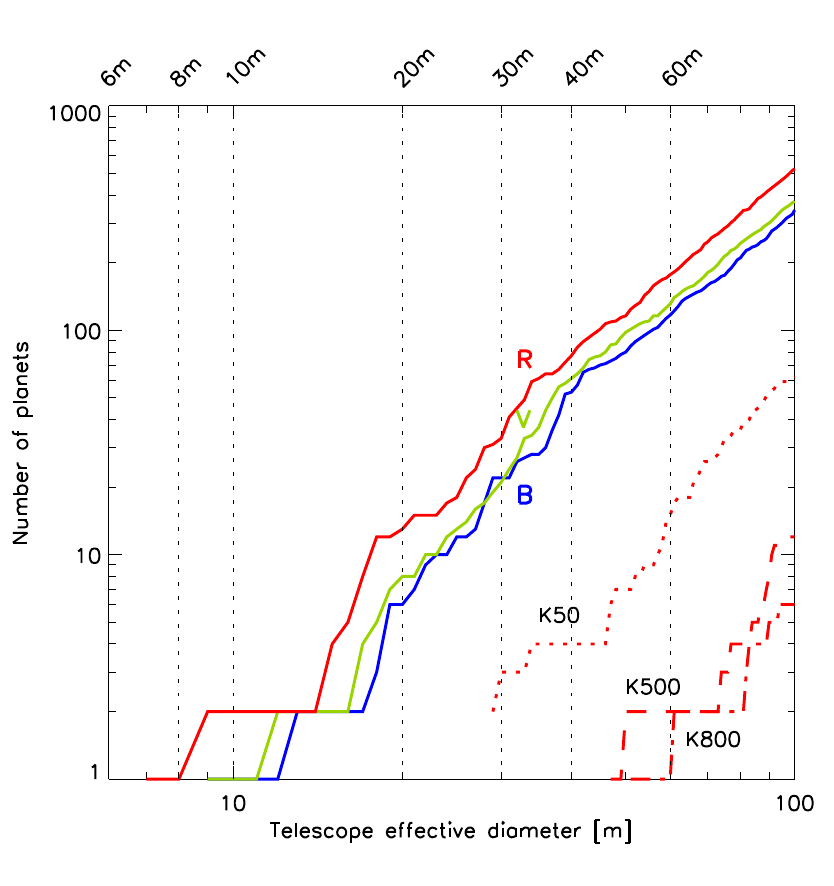}
   }
\caption{Number of detectable WHZ Earth-size planets (1\,$R_{\rm E}$) in the BVR bands (SNR$\ge$5) 
around AFGKM main sequence stars brighter than $V=13$ magnitude.
Solid lines are for unsegmented or interferometric telescopes
which can achieve the contrast as shown in Fig.~\ref{fig:psf}.
Dotted, dashed and dashed-dotted lines are for Keck-type telescopes
(segmented primary and single secondary mirrors) with the number of segments
of 50, 500, and 800, respectively.
Here, the exposure time is 4h, the planet albedo is 0.2 in all bands, illumination phase is 0.5. 
}
\label{fig:npl}
\end{figure}

The number of Earth-size (1\,$R_{\rm E}$) detectable planets is plotted for different telescope sizes
and types in Fig.~\ref{fig:npl}. 
The plot shows that a 20m-class telescope with the SPHERE-like performance
(like the ELF) can detect and surface image a dozen of WHZ Earth-like planets.
Hundreds of such planets in this magnitude-limited sample could be detected by a 70\,m
telescope (like the Colossus). 
However, we evaluate that most probably none of such planets can be detected by Keck-style 
telescopes with $D<45$\,m.

The number of detectable super-Earths up to the Neptune size (4\,$R_{\rm E}$) 
is plotted for different telescope sizes and types in Fig.~\ref{fig:npl}.
In this case, a 20m coronagraphic telescope can detect and image hundreds of planets,
while all such planets can be discovered and imaged with a 60m high-performance telescope.
Multi-segmented Keck-style telescopes, such as TMT and EELT, should be able to collect
enough photons above the stellar background for tens of such planets.
Jupiter-size planets are more realistic targets for such telescopes.

\begin{figure}
\centering
\resizebox{7cm}{!}{
   \includegraphics{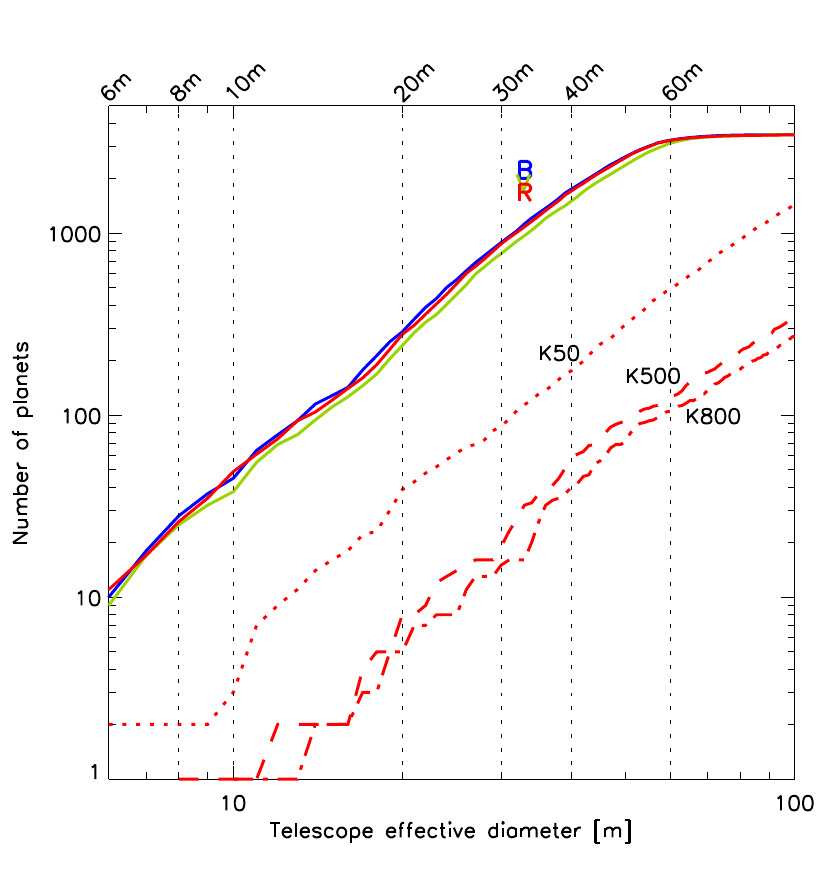}
   }
\caption{Same as Fig.~\ref{fig:npl} but for planets of 4\,$R_{\rm E}$, which is about the size of Neptune.
All such planets (and bigger ones) in the solar neighbourhood could be detected and surface imaged 
with a 70m Colossus-like telescope.
}
\label{fig:npl4}
\end{figure}

\section{Summary and Conclusions}\label{sec:con}

This paper demonstrates that for many exoplanets it is possible to map 
their surface or cloud structures using reflected starlight analyzed 
with the ExoPlanet Surface Imaging (EPSI) technique. 
We have shown that time resolved exoplanet 
photometry (on the scale of a few percent of the exoplanet rotation period) 
and with the SNR of as low as 20, can yield enough information to, for example, detect
continents on a water-rich world. With higher SNR it may be possible to trace the outline of continents
and their large-scale albedo features, such as deserts, vegetation areas, snowfields, icecaps, etc., 
or even  artifical "alien megastructures."

Combining EPSI with spectrally resolved 
data may yield information about exoplanetary subcontinental-scale
biomarkers. Planetary "noise" like clouds would limit such observations, 
but under some circumstances only by decreasing the surface albedo contrast. 

These data are within reach of the next generation of coronagraphic telescopes. 
For example a dedicated hybrid telescope-interferometer of 12--20m diameter could
generate surface maps in different colors for the nearest exoplanet, Proxima b, and a few others. 
Dozens of Earth-size exoplanets and hundreds of larger planets 
could be imaged this way with a hybrid telescope of 30m or larger diameter.

SVB acknowledges the support by the ERC Advanced Grant HotMol (www.hotmol.eu). 
JRK acknowledges the support by the Alexander von Humboldt Foundation, Germany.


\end{document}